\documentclass[conference]{IEEEtran}
\IEEEoverridecommandlockouts
\usepackage[T1]{fontenc}
\usepackage[utf8]{inputenc}
\usepackage{microtype}
\usepackage{amsmath,amssymb,amsthm}
\usepackage{mathtools}
\usepackage{xspace}
\usepackage{booktabs}
\usepackage{multirow}
\usepackage{makecell}
\usepackage{siunitx}
\usepackage[inline]{enumitem}
\usepackage{algorithm}
\usepackage{algpseudocode}
\usepackage{graphicx}
\usepackage{tikz}
\usetikzlibrary{shapes.geometric,arrows.meta,positioning,fit,backgrounds,patterns}
\usepackage{pgfplots}
\pgfplotsset{compat=1.18}
\usepackage{hyperref}
\hypersetup{hidelinks}
\usepackage{balance}
\usepackage{array}
\usepackage{colortbl}
\usepackage{tcolorbox}
\usepackage{listings}
\tcbuselibrary{listings,skins}
 
\newtheorem{definition}{Definition}

\algrenewcommand\algorithmicrequire{\textbf{Input:}}
\algrenewcommand\algorithmicensure{\textbf{Output:}}
 
\newcommand{\cape}{\textsc{Cape}\xspace}
\newcommand{\CVSS}{\textsc{cvss}\xspace}
\newcommand{\EPSS}{\textsc{epss}\xspace}
\newcommand{\KEV}{\textsc{kev}\xspace}
\newcommand{\SCA}{\textsc{sca}\xspace}
\newcommand{\SBOM}{\textsc{sbom}\xspace}
\newcommand{\CISA}{\textsc{cisa}\xspace}
\newcommand{\AHP}{\textsc{ahp}\xspace}
\newcommand{\VEX}{\textsc{vex}\xspace}
\newcommand{\NVD}{\textsc{nvd}\xspace}
\newcommand{\OSV}{\textsc{osv}\xspace}

\newcommand{\RQone}{\textbf{RQ1}\xspace}
\newcommand{\RQtwo}{\textbf{RQ2}\xspace}
\newcommand{\RQthree}{\textbf{RQ3}\xspace}
\newcommand{\RQfour}{\textbf{RQ4}\xspace}
\definecolor{capeblue}{RGB}{31,119,180}
\definecolor{tsgray}{RGB}{70,130,180}
\definecolor{pygold}{RGB}{44,160,44}
\definecolor{r2col}{RGB}{186,186,186}
\definecolor{r3col}{RGB}{253,191,111}
\definecolor{r4col}{RGB}{107,174,214}
\definecolor{r5col}{RGB}{189,55,54}
\definecolor{wincol}{RGB}{198,232,196}
\definecolor{losecol}{RGB}{252,196,192}
\definecolor{tiecol}{RGB}{255,255,204}
\definecolor{hdrcol}{RGB}{240,240,240}
\definecolor{fporange}{RGB}{255,200,120}   
\definecolor{fpblue}{RGB}{180,210,240}     
 
\def\BibTeX{{\rm B\kern-.05em{\sc i\kern-.025em b}\kern-.08em
    T\kern-.1667em\lower.7ex\hbox{E}\kern-.125emX}}

\begin{document}

\title{CAPE — Context-Aware Prioritization Engine for Vulnerability Scoring and Management}


\author{
\IEEEauthorblockN{Rakesh Podder, Viktoria Koscinski, and Indrajit Ray}
\IEEEauthorblockA{Computer Science Department, Colorado State University, Fort Collins, Colorado, USA \\ 
\{Rakesh.Podder, Viktoria.Koscinski, Indrajit.Ray\}@colostate.edu}
\\ 
}

\maketitle

\begin{abstract}
Vulnerability patching is critical for safeguarding an enterprise's software ecosystem from cyber attacks. However, patching requires adequate preparation for potential system downtime and for unforeseen effects of the patched software on a live system. Thus, effective risk mitigation requires prioritization strategies that explicitly account for the operational context of the affected software.
While vulnerability prioritization based on Common Vulnerability Scoring System (\CVSS), Exploit Prediction Scoring System (\EPSS) and  CISA Known Exploited Vulnerabilities (\KEV) catalog, can help, their differing goals and methodologies often lead to inconsistent, context-unaware decisions resulting in mis-allocated remediation efforts. In this work, we present the Context-Aware Prioritization Engine (\cape), a vulnerability analysis framework that enriches each vulnerability scanner report of vulnerabilities (\textsc{cve}s)  with deployment-specific evidence. This evidence is used to perform vulnerability reachability, centrality, and exploitability analysis, producing a ranked priority score for the effective distribution of remediation effort. We evaluate \cape on 30 open-source agentic applications comprising 6,138 scanner-reported \textsc{cve}s. Our evaluation shows that a significant majority of \textsc{cve}s are not reachable and exploitable which results in significant disparities between \cape and context-free baselines — the \CVSS, \EPSS, and \KEV. It also shows that \cape consistently surfaces confirmed-exploitable \textsc{cve}s at higher priority than all baselines, demonstrating that context-aware analysis provides substantially more accurate remediation guidance than context-free severity scoring alone.
\end{abstract}



\section{Introduction}
\label{sec:intro}

Modern software development and supply chains are undergoing a structural shift. The rise of AI-assisted coding tools, autonomous agentic frameworks, and large-language-model-powered development environments has dramatically accelerated both the pace of software production and the depth of third-party dependency consumption~\cite{france2024navigating,stray2025human,robbes2026agentic,li2025rise,podder2026hidden}. A single agentic framework, such as LangChain~\cite{langchain}, AutoGPT~\cite{autogpt}, or CrewAI~\cite{crewai}, often depends on hundreds of specialized open-source packages for tool use, memory, retrieval, and model invocation.
Unfortunately, this dependency on third-party libraries creates a compounding vulnerability management problem as dependencies cannot be freely upgraded to mitigate software vulnerabilities. Breaking changes in transitive dependencies, incompatible API revisions, and brittle build chains force development teams to pin packages at specific versions~\cite{stringer2020technical,mens2024overview,jayasuriya2025extended}. This phenomenon, commonly called \emph{dependency hell}~\cite{abate2020dependency}, often results in known-vulnerable versions remaining in production long after patches are available. A study of \textit{npm} packages found that the median time between a vulnerability disclosure and a consuming project updating the affected dependency exceeds four months~\cite{chinthanet2021lags,pashchenko2018vulnerable}. 

Software Composition Analysis (\SCA)~\cite{zhao2023software, dietrich2023security, zhan2021atvhunter, Zhuobing2017vul} can help mitigate this problem to some extent. Tools such as Syft~\cite{syft} and Trivy~\cite{trivy} generate a Software Bill of Materials (\SBOM) from an installed dependency tree and vulnerability scanners such as Grype~\cite{grype}, Trivy~\cite{trivy}, and CVE-bin-tool~\cite{cvebintool} match each component name version against the National Vulnerability Database (\NVD)~\cite{nvd}, the Open Source Vulnerability database~\cite{osv}, and ecosystem-specific databases. The output is a list of vulnerabilities (henceforth referred to as \textsc{cve}s). For a non-trivial application, this list routinely contains hundreds of entries~\cite{zhao2023software}. Recent \NVD trends reinforce the scale of the problem: 40,077 \textsc{cve}s were published in 2024 and 48,185 in 2025~\cite{cvemetrics}. 
Once the vulnerability scanner output list of \textsc{cve}s for a software is available, the dominant practice is to sort the \textsc{cve}s by \CVSS base score and address them top-down~\cite{wermke2022trust}. Alternatively, practitioners filter \textsc{cve}s by \EPSS probability~\cite{first_epss} or \CISA \KEV membership~\cite{cisa_kev}. 

We observe two critical weaknesses in this practice. \textbf{First}, \SCA scanners match on \texttt{(component,\,version)} tuples without inspecting the deployed artifact. Consequently, a build-time bundle excluded from production, a component with a backported fix lacking a version bump, and a library imported on every request produce identical scanner outputs despite having radically different operational risks. The result is a substantial fraction of reported \textsc{cve}s that cannot be triggered in any execution of the deployed application. We term these \emph{context-aware false positives}. \textbf{Second}, \CVSS base scores are assigned against generic target systems~\cite{cvss4, cvss31}, carrying no information about (i) whether the vulnerable component is architecturally central to the application, (ii) whether the buggy code is physically present in the installed artifact or has been patched out, (iii) whether the dependency is direct or transitive, or (iv) whether any execution path from an application entry point reaches the vulnerable code. Two \textsc{cve}s with identical \CVSS scores can represent completely different operational risks. Sorting hundreds of alerts purely by \CVSS forces developers, who already face intense pressure to prioritize features over security~\cite{assal2018security}, to allocate effort by noise rather than actual risk. The financial cost of this misallocation is substantial~\cite{assal2018security,bandara2021large}. One survey of 600 cybersecurity professionals found that 63\% are unable to act on alert volume and 67\% lack the time to mitigate all vulnerabilities~\cite{Ponemon2020VulnMgmt}.

Both of these observed weaknesses share a root cause: traditional \SCA frameworks are \emph{context-blind}. Prior work has proposed context-aware prioritization, but existing systems draw context from organizational history, infrastructure state, or service topology. None inspect the deployed artifact at the source-code level~\cite{cavp2022,ahmadi2022automated,jiang2012vrank,siewruk2021context,huang2026aegisguard}. \textsc{cavp}~\cite{cavp2022} and \textsc{acvrm}~\cite{ahmadi2022automated} adjust scores via expert heuristics and organizational criteria respectively; \textsc{vrank}~\cite{jiang2012vrank} scores by inter-service role in composed \textsc{soa} systems; AegisGuard~\cite{huang2026aegisguard} infers applicability from host telemetry and retrieval-augmented threat intelligence. Static reachability has been studied for Java and Maven ecosystems~\cite{soto2021bloat,dann2021identifying}, finding that only 11--13\% of vulnerabilities are method-reachable in dependent projects, but neither work extends reachability to prioritization or patch-state confirmation. No prior system performs the conjunction of entry-point reachability analysis, artifact-level bug or patch-line signature matching, and graph-theoretic centrality measurement within a single, formally specified scoring model.

We present the Context-Aware Prioritization Engine (\cape), an analysis framework that addresses both of these problems by enriching each \textsc{cve} finding with deployment-specific evidence before scoring it. \cape constructs a dependency graph of the target system with heterogeneous programming languages (python, java, js/ts, go, ruby, etc.) as an abstract syntax tree (\textsc{ast}). Multi-source breadth-first reachability analysis from application entry points eliminates context-aware false positives rather than merely deprioritising them. Fix GitHub commits retrieved from \NVD and \OSV are matched against installed artifact line signatures, confirming whether buggy or patched code is physically present regardless of declared version. Graph-theoretic centrality metrics — impact radius, hub score, and eigenvector centrality — quantify structural influence of affected components and files with bugs beyond what reachability alone expresses. This evidence is combined in a formally specified multi-dimensional scoring model with Analytic Hierarchy Process weights~\cite{saaty1980ahp}, weight redistribution for absent dimensions, Vulnerability Exploitability eXchange (VEX~\cite{openvex}) style regimes, and gate logic to create a ranked prioritization schema of \textsc{cve}s that a developer can act on.

\cape is evaluated on 30 open-source projects spanning Python, TypeScript, and JavaScript, comprising 6,138 scanner-reported \textsc{cve}s. We address four research questions.

\begin{description}[leftmargin =0pt, itemsep=2pt]
  \item[\RQone] \textit{What fraction of scanner-reported \textsc{cve}s are statically unreachable or exploitable from any application entry point, and how does this rate vary across ecosystems?}
  \item[\RQtwo] \textit{Does \cape prioritize confirmed-exploitable \textsc{cve}s more effectively than \CVSS, \EPSS, and \KEV baselines, and under what structural conditions does each baseline compete?}
  \item[\RQthree] \textit{How consistently does \cape separate high-evidence \textsc{cve}s from low-evidence \textsc{cve}s relative to \CVSS, \EPSS, \EPSS-percentile, and \KEV baselines?}
  \item[\RQfour] \textit{How sensitive is \cape's ranking to perturbations of the \AHP weight prior, and how does the choice of centrality metric selection affect ranking quality across cases?}
\end{description}

The primary contributions of this work are:

\begin{itemize}

\item We present Context-Aware Prioritization Engine (\cape), a vulnerability analysis framework that enriches each vulnerability scanner-reported \textsc{cve} with static reachability, code-level evidence, and graph-theoretic centrality into a context-aware priority ranking.

\item We introduce a regime-adaptive \textsc{ahp} scoring model that marginalizes structurally absent evidence dimensions, preserving score monotonicity across deployment-specific evidence and VEX verdicts for each \textsc{cve}.

\item We quantify scanner-reported \textsc{cve}s that are statically unreachable at scale. 43.3\% of 6{,}138 scanner-reported \textsc{cve}s across 30 open-source projects are statically unreachable, reaching 76.1\% in the most extreme case and averaging 53.2\% for TypeScript/JavaScript projects.

\item We evaluate \cape against four baselines, \CVSS, \KEV, \EPSS, and \EPSS percentile,  on 30 open-source applications, achieving a 79.8\% reduction in mean rank of first confirmed-exploitable \textsc{cve} (3.4 vs.\ 16.9; median 1 vs.\ 7), with \textsc{auroc} 0.908 vs.\ 0.454 and \textsc{ndcg}@10 0.652 vs.\ 0.227 for \textsc{cvss}.

\end{itemize}
All code, data, and evaluation artifacts will be available upon publication. 


\section{Background \& Related Work}
\label{sec:background}

In the field of vulnerability research and practice, the use of modern \SCA tools~\cite{syft,grype,trivy,cvebintool} are prevalent for detecting \textsc{cve}s in software and \CVSS scores~\cite{zhao2023software,dietrich2023security} for remediation decision-making. Such vulnerability scanner output contains hundreds of \textsc{cve}s~\cite{zhao2023software}, yet practitioners are unable to act on the alert volume~\cite{koscinski2025conflicting}. \CVSS Base scores~\cite{cvss4, cvss31} are the dominant choice for mitigation recommendation regardless of reachability, code evidence, or architectural centrality. 
\EPSS~\cite{jacobs2021epss,jacobs2023enhancing} is another scoring method which gives 30-day exploitation predictions but clusters near zero for recently published \textsc{cve}s. Suciu et al.~\cite{suciu2022expected} introduce a time-varying Expected Exploitability metric that partially mitigates \EPSS's recency gap but remains deployment-agnostic. \KEV~\cite{cisa_kev} provides empirical exploitation confirmation but covers only $\sim$1.4\% of all \textsc{cve}s. Koscinski et al.~\cite{koscinski2025conflicting} compare \CVSS, \EPSS, \textsc{ssvc}, and the Exploitability Index across 600 real-world vulnerabilities, finding significant rank disparities and no single system reliably capturing exploitation risk. Hybrid scoring proposals~\cite{sharma2023hybrid,shimizu2026vulnerability,albanese2023framework,farris2018vulcon} adjust weighting schemes, reducing remediation workload~\cite{shimizu2026vulnerability}, but none incorporate deployment-specific source-code evidence.

Recognizing generic scores as insufficient, researchers introduced context from three sources, organizational history, infrastructure state, and service topology, but none investigated the deployed artifact or software~\cite{cavp2022,ahmadi2022automated,jiang2012vrank,siewruk2021context,huang2026aegisguard,jiang2025survey}. 
\textsc{cavp}~\cite{cavp2022}, \textsc{acvrm}~\cite{ahmadi2022automated}, and \textsc{ssvc}~\cite{spring2021prioritizing} refine severity scoring via expert heuristics or organizational context but perform no source-code analysis. \textsc{vrank}~\cite{jiang2012vrank} and ChainFuzz~\cite{deng2025chainfuzz} incorporate structural context but are scoped to \textsc{soa} topologies and proof-of-concept reachability, respectively. Mixeway~\cite{siewruk2021context} and AegisGuard~\cite{huang2026aegisguard} use scan history and host telemetry as context, neither of which inspects the installed artifact. Table~\ref{tab:related} shows that no prior system combines entry-point reachability, artifact-level code evidence, and component centrality.

\begin{table}[t]
\centering
\tiny
\caption{Comparison of vulnerability prioritization approaches. \checkmark~=~fully addressed; $\circ$~=~partially addressed;
\texttimes~=~not addressed.}
\label{tab:related}
\resizebox{\columnwidth}{!}{%
\begin{tabular}{lp{1.8cm}ccccc}
\toprule
\textbf{System} & \textbf{Context Source} &
\textbf{Entry-Point} & \textbf{Artifact} & \textbf{Component} &
\textbf{Dep.} & \textbf{Multi-Lang} \\
 & & \textbf{Reach.} & \textbf{Code Evidence} & \textbf{Centrality} &
\textbf{Type} & \textbf{Software}\\
\midrule
CVSS~\cite{cvss4}          & None (generic)         & \texttimes & \texttimes & \texttimes & \texttimes & \texttimes \\
EPSS~\cite{jacobs2021epss} & Threat intel           & \texttimes & \texttimes & \texttimes & \texttimes & \texttimes \\
KEV~\cite{cisa_kev}        & Exploit confirmation   & \texttimes & \texttimes & \texttimes & \texttimes & \texttimes \\
SSVC~\cite{spring2021prioritizing}           & Analyst decision tree  & \texttimes & \texttimes & \texttimes & \texttimes & \texttimes \\
VIEWSS~\cite{sharma2023hybrid} & Score reweighting  & \texttimes & \texttimes & \texttimes & \texttimes & \texttimes \\
VMC~\cite{shimizu2026vulnerability}  & KEV+EPSS filter chain  & \texttimes & \texttimes & \texttimes & \texttimes & \texttimes \\
CAVP~\cite{cavp2022}       & Expert heuristics      & \texttimes & \texttimes & \texttimes & \texttimes & \texttimes \\
ACVRM~\cite{ahmadi2022automated}     & Org.\ risk appetite    & \texttimes & \texttimes & \texttimes & \texttimes & \texttimes \\
VRank~\cite{jiang2012vrank}     & SOA service topology   & $\circ$    & \texttimes & $\circ$    & \texttimes & \texttimes \\
Mixeway~\cite{siewruk2021context} & Scan history    & \texttimes & \texttimes & \texttimes & \texttimes & $\circ$ \\
AegisGuard~\cite{huang2026aegisguard} & Host telemetry  & \texttimes & \texttimes & \texttimes & \texttimes & \texttimes \\
DepClean~\cite{soto2021bloat} & Source code        & $\circ$    & \texttimes & \texttimes & \texttimes & \texttimes \\
Dann et al.~\cite{dann2021identifying} & Source code & $\circ$ & \texttimes & \texttimes & \texttimes & \texttimes \\
ChainFuzz~\cite{deng2025chainfuzz} & Source code (PoC) & \checkmark & \texttimes & \texttimes & \texttimes & \texttimes \\
\midrule
\textbf{CAPE (ours)}       & \textbf{Source code + Installed artifact + SBOM} &
\checkmark & \checkmark & \checkmark & \checkmark & \checkmark \\
\bottomrule
\end{tabular}%
}
\end{table}

Static reachability applied to \textsc{sca} for vulnerabilities has been well-explored for Java~\cite{kula2018developers,dann2021identifying,wang2020empirical} and Maven~\cite{soto2021bloat,ma2024vulnet} ecosystems. Soto-Valero et al.~\cite{soto2021bloat} find that 75\% of Maven dependencies contain at least one unreachable method, while Dann et al.~\cite{dann2021identifying} report only 11--13\% of vulnerabilities are method-reachable in dependent projects; neither extends this to prioritization. ChainFuzz~\cite{deng2025chainfuzz} validates upstream vulnerabilities via \textsc{poc} generation in C/C++ supply chains, confirming reachability but not generalizing to cross-language prioritization pipelines. For code evidence, Sabetta and Bezzi~\cite{sabetta2018patch} show that patch commits carry distinctive code signatures. Li and Paxson~\cite{li2017large} and Bhandari et al.~\cite{bhandari2021cvefixes} establish at scale that downstream dependents remain unpatched long after fixes are published~\cite{chinthanet2021lags,pashchenko2018vulnerable,liao2020measuring}. Bandara et al.~\cite{bandara2021large} show that 18\% of all vulnerabilities in open-source \textsc{js} projects originate from git commits, making version-number-based patch status an unreliable signal. For dependency graph construction, static analysis of dynamic languages involves known precision-recall trade-offs~\cite{feldthaus2013callgraph,madsen2013javascript,ali2014averroes}. 
The \VEX~\cite{openvex} standard encodes producer-asserted exploitability status, but fewer than 5\% of advisories carry \VEX statements~\cite{cisa_kev} and producer assertions do not account for deployment context. \cape automates \VEX-equivalent evidence from the consumer side. \cape combines entry-point BFS reachability analysis, code evidence with buggy \& patch-code matching, and graph-theoretic centrality measures into a regime (VEX type)-adaptive scoring model with evaluation against state-of-the-art baseline scores.


\section{Formal Problem Statement}
\label{sec:problem}

Let $\Sigma$ denote a software system under analysis, characterized by its source repository and dependency manifest. \cape generates a deployment-representative \SBOM (using Syft) via local build, modeled as $\mathbb{B} = \beta(\Sigma)$ --- a machine-readable inventory of every component at its exact resolved version (\S~\ref{sec:pipeline}). A vulnerability scanner (Grype) applied to $\mathbb{B}$ produces \textsc{cve} \texttt{($\mathsf{pkg}=$component,\,$\mathsf{ver}=$version,\,$\mathsf{cve}=$\textsc{cve})} list.

\begin{definition}
\label{def:finding}
A \emph{vulnerability or \textsc{cve} list} is a triple $f = (\mathsf{cve},\, \mathsf{pkg},\, \mathsf{ver}) \in \mathcal{C} \times \mathcal{P} \times \mathcal{V}$, where $\mathcal{C}$, $\mathcal{P}$, $\mathcal{V}$ are the universes of \textsc{cve} identifiers, package names, and version strings. The scanner applied to $\beta(\Sigma)$ produces a finite \textsc{cve} set $\mathcal{F}_\mathbb{B} = \{f_1, \ldots, f_n\}$.
\end{definition}


Each $f \in \mathcal{F}_\mathbb{B}$ is scanner output only. It carries no information about whether the vulnerable code is reachable in $\Sigma$, whether the installed artifact incorporates the fix, or how architecturally central the component is.
Across 30 evaluated projects, $|\mathcal{F}|$ ranges from 8 to 719, totalling 6,138 \textsc{cve}s. \textbf{43.3\%} are statically unreachable (TypeScript/JavaScript: 53.0\%; Python: 27.4\%) and constitute \emph{context-aware false positives} \textsc{cve}s that cannot be triggered in any execution of the deployed application yet are indistinguishable from exploitable ones in raw scanner output.
This context-blindness motivates a priority function that ranks $\mathcal{F}_\mathbb{B}$ by operational exploitability rather than by generic severity.



\begin{definition}
\label{def:priority}
A \emph{priority function} $\mathbb{P} : \mathcal{F}_\mathbb{B} \to [0,1]$ inducing a total preorder $f_a \succeq f_b \text{ iff } \mathbb{P}(f_a) \geq \mathbb{P}(f_b)$, approximating the security preference ``remediate $f_a$ before $f_b$.''
\end{definition}

The current practice uses the $\mathsf {CVSS}(f)$ priority 
which fails on two counts. First, a reachable confirmed-bugged vulnerability may score lower than an unreachable one, and second, all deployment-specific evidence is ignored. 


\section{The \cape Architecture}
\label{sec:pipeline}



\begin{figure*}[t]
    \centering
    \includegraphics[width=0.8\linewidth]{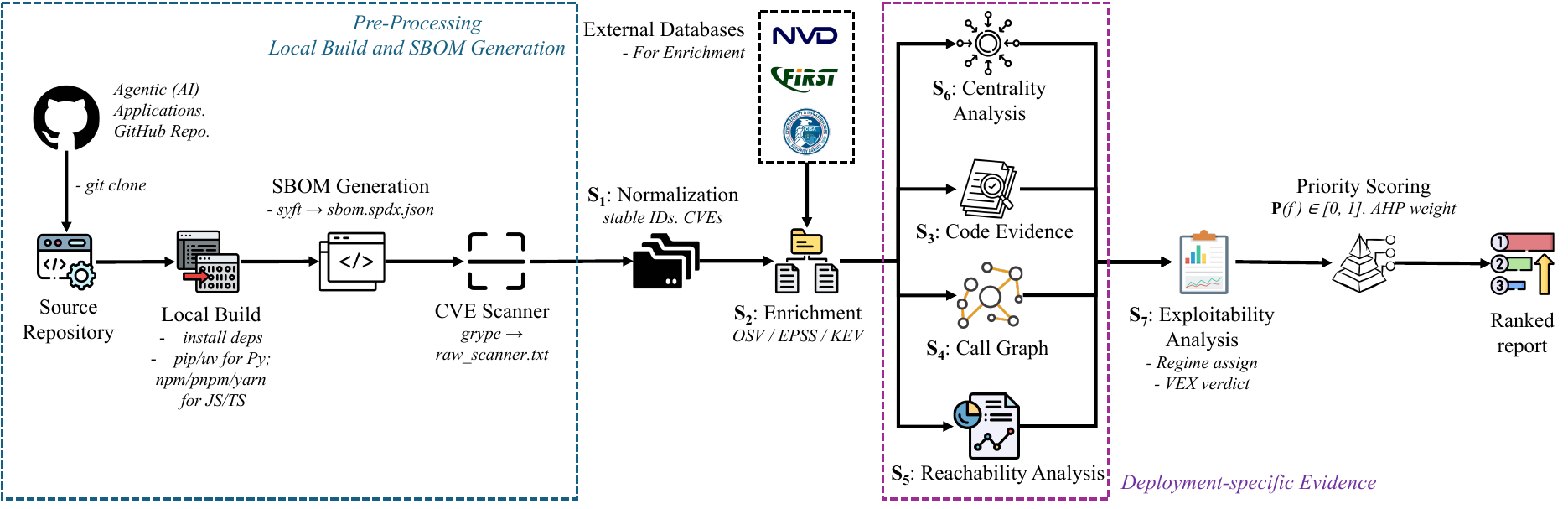}
    \caption{\cape Architecture. Stages $\mathcal{S}_3$--$\mathcal{S}_6$ are mutually independent and execute in parallel; $\mathcal{S}_7$ requires all 4 stage signals or data;  scoring model requires $\mathcal{S}_7$.}
    \label{fig:architecture}
\end{figure*}

\cape transforms scanner \textsc{CVE}s into a context-aware priority ranking through evidence collected from seven stages $\Pi = (\mathcal{S}_1, \ldots, \mathcal{S}_7)$, preceded by a pre-processing phase that generates a deployment-representative \SBOM via local build (\S~\ref{sec:localbuild}). Figure~\ref{fig:architecture} shows the complete architecture. Throughout \S~\ref{sec:pipeline} and \S~\ref{sec:model}, we use OpenClaw~\cite{repo-openclaw} as a running example to illustrate how evidence flows through each stage of the CAPE pipeline

$\mathcal{S}_1$--$\mathcal{S}_2$ normalize and enrich each \textsc{cve} with \OSV metadata, \EPSS probabilities, and \KEV membership. $\mathcal{S}_3$--$\mathcal{S}_6$ collect artifact-level code-evidence signature matching against GitHub commits ($\mathcal{S}_3$), a typed directed call graph via modified Tree-sitter ($\mathcal{S}_4$), multi-source \textsc{bfs} reachability from application entry points ($\mathcal{S}_5$), and graph-theoretic centrality of affected components and files ($\mathcal{S}_6$). $\mathcal{S}_7$ synthesizes these streams into a deterministic exploitability verdict and regime $k(f) \in \{\mathcal{R}_0,\ldots, \mathcal{R}_5\}$. The scoring model (\S~\ref{sec:model}) then produces $\mathbb{P}(f) \in [0,1]$ based on these evidence per \textsc{cve}.

\subsection{Local Build and \SBOM Generation}
\label{sec:localbuild}






Platform-generated \SBOM{}s (e.g., GitHub Dependency Graph~\cite{githubdependencygraph}) are  unsuitable for vulnerability analysis. We found that GitHub-generated \SBOM{}s or strong project manager (PM)-based \SBOM{}s \cite{zhou2025reality} cannot observe the deployment context, producing version divergence, architecture-dependent false positives (e.g., all 23 \texttt{@esbuild} platform variants listed instead of the one installed), and staleness from repository churn. \cape{}  generates $\mathbb{B} = \beta(\Sigma)$ locally via 
\textit{(i)~source acquisition} -- cloning the target repository at a pinned commit into an isolated case directory, ensuring the \SBOM{}, scanner output, and call graph all correspond to the same code state; \textit{(ii)~dependency installation} -- installed using the ecosystem's native toolchain, 
selected automatically from the lockfile, with post-install hooks disabled to prevent arbitrary code execution; \textit{(iii)~\SBOM{} generation} -- \texttt{syft} scans the installed dependency tree (\texttt{.venv/} or
\texttt{node\_modules/}) to produce a standardized \textsc{spdx}-2.3 file at exact resolved versions. 

We use \texttt{grype}~\cite{grype} 
against $\mathbb{B}$ to get a \textsc{CVE} list containing one row per \texttt{(component, version, \textsc{cve})} triple into \texttt{raw\_scanner.csv} for $\mathcal{S}_1$. Every result in this paper is computed against locally built \SBOM{}s.

\smallskip
\noindent\fbox{\parbox{0.96\columnwidth}{\small
\textbf{Observation.} In \texttt{eslint}, \texttt{@babel/code-frame} resolved to \texttt{7.10.1} on GitHub vs.\ \texttt{7.0.0} locally -- different transitive trees, different \textsc{cve}s. In \texttt{OpenClaw}, the platform listed all 23 \texttt{@esbuild} variants; local build installed one, making 22 findings unreachable false positives. Lock-file-only installs produced version divergence across 6 of 12 \texttt{npm} projects, confirming neither source reliably reflects the deployed artifact.
}}

\subsection{Stages \texorpdfstring{$\mathcal{S}_1$--$\mathcal{S}_2$}
{S1-S2}: Normalisation and Enrichment}
\label{sec:s1s2}

\noindent\textbf{$\mathcal{S}_1$ --- Normalization.} Raw scanner output is a multiset of rows where a single \textsc{cve} may appear under multiple advisory aliases (\textsc{nvd}, \textsc{ghsa}, \textsc{osv}). $\mathcal{S}_1$ duplicates on the identity projection $\pi(r) = (r[\mathsf{pkg}],\, r[\mathsf{ver}],\, r[\mathsf{cve}])$, collapsing aliases while keeping distinct \textsc{cve}s on the same \texttt{(pkg,~ver)} pair separate.
The output is an ordered \textsc{cve} set $\mathcal{F} = (f_1, \ldots, f_n)$ with stable sequential identifiers for the downstream process.

\smallskip
\noindent\textbf{$\mathcal{S}_2$ --- Enrichment.} Each $f \in \mathcal{F}$ is augmented with three external signals queried live at execution time: \textit{(i)~advisory metadata} --- the \OSV \textsc{api} is queried per \textsc{cve}, when found, summary and affected version ranges are extracted;
\textit{(ii)~\EPSS probability}~\cite{jacobs2021epss} --- a daily-updated 30-day exploitation forecast stored in $f.\mathsf{epss}$, used as an external baseline only; and \textit{(iii)~\KEV}~\cite{cisa_kev} --- a Boolean flag $f.\mathsf{kev}$ confirming wild exploitation, also used as an external baseline for evaluation (\S~\ref{sec:eval}) only. Please not that, we are considering \CVSS score as external severity signal in our evidence dimension (\S~\ref{sec:dimensions}) for it's consistency and wide adaptation as industry standard.

\noindent\textit{OpenClaw.} Table~\ref{tab:S1-S7} illustrates enrichment output for OpenClaw \textsc{cve}s. As can be observed from the table, the highest-CVSS entry (f\_003, CVE-2026-33937, score~9.8) sits at the 66.4th \EPSS percentile, while f\_001 (\texttt{vite 7.3.1}, CVE-2026-39363, \textsc{cvss}~8.2) reaches the 80.1 percentile with 0.02292 \EPSS score yet has no \OSV record --- neither signal nor scoring method alone is sufficient for remediation strategies.

\begin{table*}[t]
\centering
\caption{ \cape Traces for evidence in OpenClaw Repo (case\_003, $n{=}161$ \textsc{cve}s).
}
\label{tab:S1-S7}
\tiny
\setlength{\tabcolsep}{2pt}
\begin{tabular}{@{}c|lp{0.8cm}|rrrcc|ccp{2.5cm}|cp{3.0cm}|cccc|r@{}}
\toprule
& & &
\multicolumn{5}{c|}{\textbf{$\mathcal{S}_1$--$\mathcal{S}_2$}}
& \multicolumn{3}{c|}{\textbf{$\mathcal{S}_3$}}
& \multicolumn{2}{c|}{\textbf{$\mathcal{S}_5$}}
& \multicolumn{4}{c|}{\textbf{$\mathcal{S}_6$}}
& \multicolumn{1}{c}{\textbf{$\mathcal{S}_7$}} \\
\textbf{$f$} & \textbf{CVE} & \textbf{Comp./Ver.}
  & \textbf{CVSS} & \textbf{EPSS} & \textbf{EPSS-pct} & \textbf{KEV} & \textbf{OSV}
  & \textbf{Comm.} & \textbf{Verd.} & \textbf{Code Evidence}
  & \textbf{Reach.} & \textbf{BFS Path}
  & \textbf{ir$_{m}$} & \textbf{ev$_{m}$} & \textbf{hub$_{m}$} & \textbf{Lv.}
  & \textbf{Regime}  \\
\midrule
003 & CVE-2026-33937 & handlebars 4.7.8 & 9.8 & 0.0129 & 66.4\% & $\times$ & $\checkmark$ & $\checkmark$ & BG & \texttt{ javascript-compiler.js}, \texttt{precompiler.js}, \texttt{runtime.js}, $\ldots$ & $\checkmark$ & \texttt{OpenClaw$\to$handlebars} & 1.000 & 0.0013 & 0.687 & fmax & $\mathcal{R}_5$ \\
\midrule
018 & CVE-2026-26278 & fast-xml-parser 5.3.4
  & 7.5 & 0.0059 & 43.6\% & $\times$ & $\checkmark$
  & $\checkmark$ & BG & \texttt{OrderedObjParser.js},
    \texttt{DocTypeReader.js}, \texttt{OptionsBuilder.js}
  & $\times$ & --- (No Reachable Path)
  & 0.000 & 0.0004 & 0.540 & fmax
  & $\mathcal{R}_5$\\
\midrule
019 & CVE-2026-33036 & fast-xml-parser 5.3.4
  & 7.5 & 0.0058 & 43.0\% & $\times$ & $\checkmark$
  & $\checkmark$ & NB & \texttt{OrderedObjParser.js}
  & $\times$ & --- (No Reachable Path)
  & 0.000 & 0.0001 & 0.540 & fmax
  & $\mathcal{R}_4$  \\
\midrule
007 & CVE-2026-4800 & lodash 4.17.23
  & 8.1 & 0.0103 & 59.1\% & $\times$ & $\times$
  & $\checkmark$ & NB & \texttt{lodash.js} (patched; buggy lines not matched)
  & $\checkmark$ & \texttt{OpenClaw$\to$lodash}
  & 1.000 & 0.0001 & 0.0001 & fmax
  & $\mathcal{R}_4$ \\
\midrule
029 & CVE-2026-2229 & undici 7.22.0
  & 7.5 & 0.0049 & 38.2\% & $\times$ & $\checkmark$
  & $\times$ & NC & --- (no git commit retrieved)
  & $\checkmark$ & \texttt{OpenClaw$\to$undici}
  & 1.000 & 0.0001 & 0.000 & comp
  & $\mathcal{R}_2$  \\
\bottomrule
\end{tabular}
\end{table*}
\subsection{Stage \texorpdfstring{$\mathcal{S}_3$}{S3}: 
Code Evidence Collection}\label{sec:s3}

Standard vulnerability (Gyrpe, Trivy) scanners match on \texttt{(component,~version)} tuples but cannot determine whether vulnerable code is physically present in the installed artifact. 
$\mathcal{S}_3$ resolves this by fetching the git commit for each \textsc{cve} and searching the installed artifact for pre-fix (bug) and post-fix (patch) line signatures. For $f = (\mathsf{cve}, \mathsf{pkg}, \mathsf{ver})$, let $\mathbb{D} = \{\psi_1, \ldots, \psi_p\}$ be the unified diff of the fix commit decomposed into file blocks. Each block $\psi_i$ yields the \emph{buggy line set} $\mathrm{Buggy}(\psi_i) = \{(s,\ell) : (\texttt{-},s,\ell) \in \psi_i,\; |s| \geq \theta,\; \lnot\,\mathrm{trivial}(s)\}$ --- lines \emph{removed} by the fix and the symmetric \emph{patched line set} $\mathrm{Patched}(\psi_i)$ comprises lines \emph{added}. 
Note, bug evidence alone cannot distinguish a backported fix from an inconclusive case, and patch evidence alone cannot confirm exploitability when a fix touches unrelated files.
$\mathcal{S}_3$ queries \NVD~2.0 and \OSV for fix-commit \textsc{url}s, prioritising \NVD Patch-tagged \textsc{url}s: $U^\star(f) = U^{\mathsf{Patch}}_{\mathsf{NVD}}(f) \;\|\; U^{\mathsf{Other}}_{\mathsf{NVD}}(f) \;\|\; U_{\mathsf{OSV}}(f)$,
with duplicates removed at first occurrence. 

\noindent\textit{Verdict \& Regime Mapping.}
$\mathcal{S}_3$ assigns a verdict per \textsc{cve} based on code evidence. \textsc{bg} (affected files and buggy codes present), \textsc{nb} (affected files exist but no buggy code or patch incorporated), \textsc{nf} (neither affected files nor buggy code found in installation), or \textsc{nc} (no git commit found or \emph{inconclusive}). The worst-case rule $\mathsf{BG} \succ \mathsf{NB} \succ \mathsf{NF} \succ \mathsf{NC}$ applies across all affected files; a single confirmed-buggy file overrides all others, as presence of buggy code has higher chances of exploitation. 
Please note that not all \textsc{cve}s have a git commit. In that case, $\mathcal{S}_3$ assigns verdict \textsc{nc}, $\varphi_3$ is structurally absent (\textsf{NaN}), and its weight redistributes proportionally to the remaining four dimensions (\S~\ref{sec:weights}). The \textsc{cve} proceeds to $\mathcal{S}_5$--$\mathcal{S}_7$ scored on available evidence alone (\S~\ref{sec:s7}) -- nothing is discarded.







\begin{figure}[!h]
\centering\tiny\ttfamily
\setlength{\tabcolsep}{0pt}
\noindent\textbf{\normalfont Git Commit for CVE-2026-33937} (handlebars 4.7.8, commit 68d8df5a)\\
\noindent\textcolor{gray}{@@ lib/handlebars/runtime.js @@ resolvePartial}\\
\noindent\textcolor{red!80}{-\ 354\ \ }partial = options.partials[options.name];\\
\noindent\textcolor{green!60!black}{+\ 355\ \ }partial = lookupOwnProperty(options.partials, options.name);\\
\smallskip
\noindent\textbf{\normalfont Buggy line confirmed in OpenClaw $\cdot$ 1 install $\cdot$ patched line absent}\\
\noindent\textcolor{red!70}{$\bullet$}\ node\_modules/handlebars/lib/handlebars/runtime.js: line 354\\
\noindent\textcolor{red!70}{$\bullet$}\ node\_modules/handlebars/lib/handlebars/compiler/javascript-compiler.js: line 691
\caption{$\mathcal{S}_3$ code evidence for CVE-2026-33937 (\texttt{handlebars} 4.7.8).}
\label{fig:code-evidence}
\end{figure}


\noindent\textit{OpenClaw.} Table~\ref{tab:S1-S7} illustrates representative $\mathcal{S}_3$ outcomes in OpenClaw spanning verdicts \textsc{bg}, \textsc{nb}, and \textsc{nc}. \autoref{fig:code-evidence} shows a confirmed-buggy line that survives in the installed artifact: in \texttt{handlebars} 4.7.8 (CVE-2026-33937), the pre-fix partial lookup \texttt{options.partials[options.name]} (\texttt{runtime.js}) is present without its patched \texttt{lookupOwnProperty(\dots)} guard. The buggy code is matched in the single installed copy under \texttt{node\_modules/handlebars}. 

\subsection{Stage \texorpdfstring{$\mathcal{S}_4$}{S4}: Call Graph Construction}
\label{sec:s4}

$\mathcal{S}_4$ constructs a typed directed multi-graph of $\Sigma$ using modified Tree-sitter~\cite{tree-sitter} for abstract syntax tree (\textsc{ast}) extraction.
A \textbf{Call Graph} is a tuple $\mathcal{M} = \langle V, E, \tau_V, \tau_E, \alpha \rangle$, where $V$ carries type labels $\tau_V(v) \in \{$\texttt{file, module, function, method, class, component, external\_module}$\}$, $E$ is a multi-set of typed directed edges with $\mathcal{T}_E \supseteq \{$\texttt{CALLS, IMPORTS,
DEPENDS\_ON, CONTAINS, DECORATES, EXTENDS, IMPLEMENTS, INSTANTIATES, RE\_ EXPORTS}$\}$, and $\alpha$ assigns $\mathsf{call\_type} \in \{\mathsf{static}, \mathsf{dynamic}\}$, source line, and confidence $\in [0,1]$ per edge.

Algorithm~\ref{alg:callgraph} proceeds in three passes: internal source parsing (lines 2--6), external package parsing (lines
7--15), and import resolution (lines 16--19). Unresolved imports are rewritten via a three-level cascade such as module index lookup, specifier resolution (with \texttt{tsconfig.json} alias support), and validated stub creation, dropping any target that fails an identifier validity check to prevent dynamic expressions from generating invalid nodes. The resolution pass runs in $O(|E|)$. Parsing is $\Theta\!\bigl(\sum_{f} |f|\bigr)$, linear in source bytes and overall construction is $O(|V| + |E|)$, excluding \textsc{i/o}.

\begin{algorithm}[t]
\caption{Call graph construction ($\mathcal{S}_4$)}
\label{alg:callgraph}
\begin{algorithmic}[1]
\footnotesize
\Require repository root $\rho$, language $\lambda$
\Ensure typed multigraph $G = (V, E)$
\State $V, E \leftarrow \emptyset$
\State $\mathcal{S}_{\mathsf{int}} \leftarrow
  \textsc{CollectInternal}(\rho)$
\For{$f \in \mathcal{S}_{\mathsf{int}}$}
  \State $(V_f, E_f) \leftarrow
    \textsc{Parse}(f, \lambda)$
  \State $V \leftarrow V \cup V_f;\; E \leftarrow E \cup E_f$
\EndFor
\State $E \leftarrow E \cup
  \textsc{EmitDependsOn}(\textsc{Manifest}(\rho))$
\For{$p \in \textsc{EnumPackages}(\textsc{Externals}(\rho,\lambda))$}
  \For{$f \in \textsc{Files}(p)$}
    \State $(V_f, E_f) \leftarrow
      \textsc{Parse}(f, \lambda,
      \mathsf{scope}{=}\mathsf{ext}, \mathsf{pkg}{=}p)$
    \State $V \leftarrow V \cup V_f;\; E \leftarrow E \cup E_f$
  \EndFor
  \State $E \leftarrow E \cup
    \{(p,f,\mathsf{CONTAINS}) : f \in \textsc{Files}(p)\}$
\EndFor
\For{$(u, t, \tau) \in E$ \textbf{ with }
     $t = \langle\mathsf{unresolved}\rangle$}
  \State $t' \leftarrow \textsc{Resolve}(u, t, V)$
  \State \textbf{if} $t'$ valid: replace $(u,t,\tau)$ with
         $(u,t',\tau)$ \textbf{else}: drop
\EndFor
\State \Return $G = (V, E)$
\end{algorithmic}
\end{algorithm}

\begin{figure}[t]
\centering
\begin{minipage}{0.49\linewidth}
  \centering
  \includegraphics[width=\linewidth]{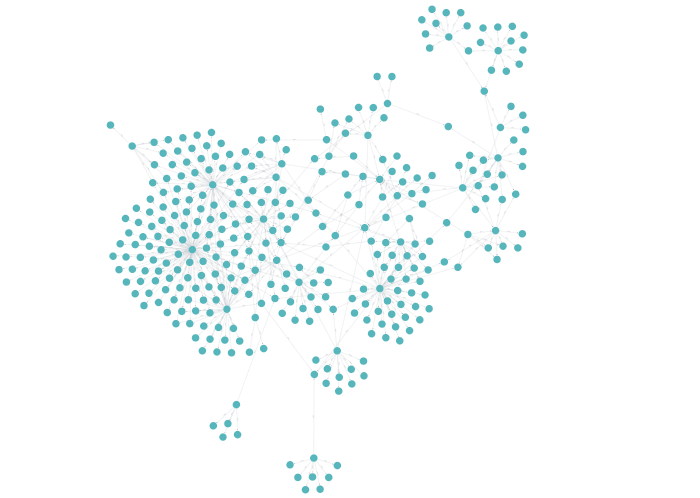}\\
  \small(a) \texttt{CALLS}: 337 nodes, 500 edges
\end{minipage}\hfill
\begin{minipage}{0.49\linewidth}
  \centering
  \includegraphics[width=\linewidth]{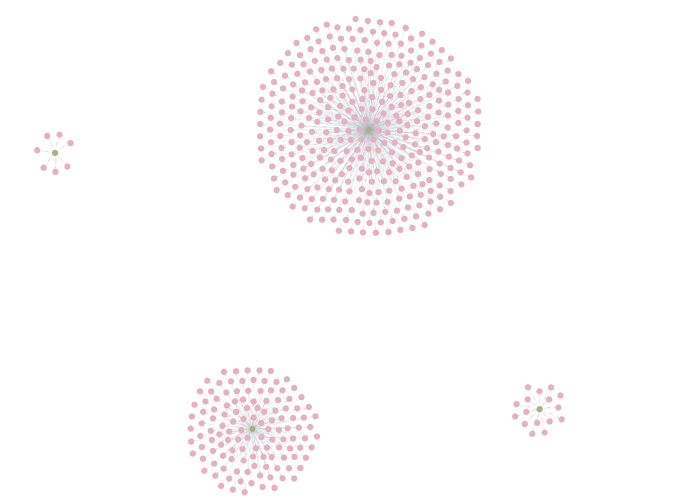}\\
  \small(b) \texttt{CONTAINS}: 504 nodes, 500 edges
\end{minipage}
\vspace{6pt}
\begin{minipage}{0.49\linewidth}
  \centering
  \includegraphics[width=\linewidth]{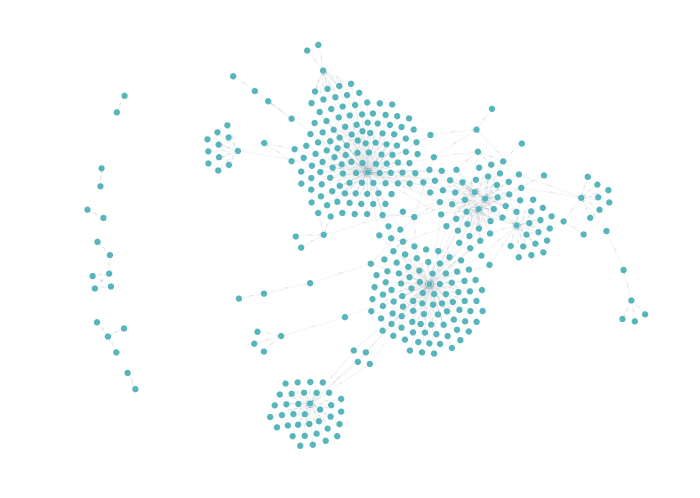}\\
  \small(c) \texttt{IMPORTS}: 392 nodes, 500 edges
\end{minipage}\hfill
\begin{minipage}{0.49\linewidth}
  \centering
  \includegraphics[width=\linewidth]{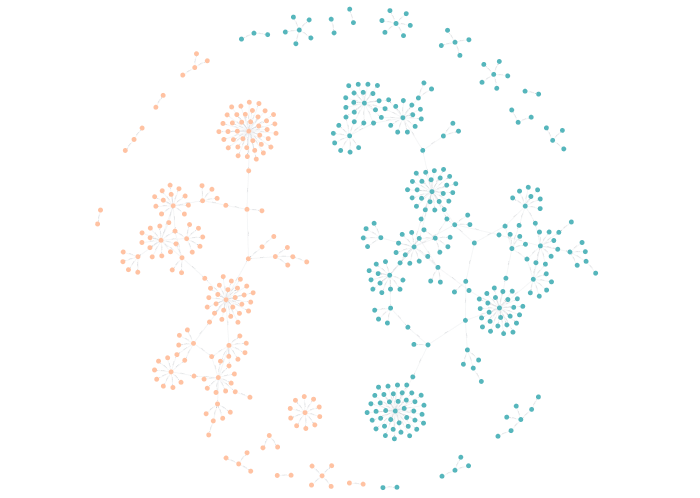}\\
  \small(d) \texttt{DEPENDS\_ON}: 489 nodes, 500 edges
\end{minipage}
\caption{Representative edge relations in the OpenClaw call graph (500-edge sample per relation).
\textcolor[HTML]{FFD86E}{\rule{6pt}{6pt}}~\textit{Function};
\textcolor[HTML]{6DCE9E}{\rule{6pt}{6pt}}~\textit{File};
\textcolor[HTML]{68BDF6}{\rule{6pt}{6pt}}~\textit{Component};
\textcolor[HTML]{FF756E}{\rule{6pt}{6pt}}~\textit{ExternalModule}.
}
\label{fig:callgraph-edges}
\end{figure}

\smallskip\noindent\textit{OpenClaw.}
$\mathcal{S}_4$ produced 316{,}491 nodes (162{,}448 function-level, 11.3\% external), 1{,}587{,}771 typed directed edges across 9 relation types, and 1 zero-in-degree entry node, consistent with OpenClaw's monolithic orchestrator architecture. As shown in Figure~\ref{fig:callgraph-edges}, \texttt{CALLS} (934{,}891) dominates the edge distribution, followed by \texttt{IMPORTS} (328{,}135) and \texttt{CONTAINS} (280{,}832). \texttt{INSTANTIATES} (15{,}628) and \texttt{RE\_EXPORTS} (27{,}770) capture object-level and module re-export dependencies.

\subsection{Stage \texorpdfstring{$\mathcal{S}_5$}{S5}: Reachability Analysis}
\label{sec:s5}

$\mathcal{S}_5$ determines, for each \textsc{cve} $f$, whether the vulnerable component lies on any executable path from an
application entry point in $\Sigma$. A negative answer identifies a context-aware false positive, meaning the the component is not exploitable. 

\smallskip\noindent\textit{Reachability Skeleton.}
$\mathcal{S}_5$ operates on a projection of $\mathcal{M}$ that removes \texttt{CONTAINS} edges and test-file nodes such that $\hat{\mathcal{M}} = (V', E')$ where $V' = V \setminus T$ and $E' = \{(u,v) : \exists\,(u,v,\tau) \in E,\, u,v \in V',\,
\tau \neq \texttt{CONTAINS}\}$, with $T$ matching patterns \texttt{.test.}, \texttt{.spec.}, \texttt{/test/}, \texttt{/tests/}.
$\hat{\mathcal{M}}$ is the \emph{reachability skeleton} --- a control-flow projection of $\mathcal{M}$, retaining only runtime-traversable edges. \texttt{CONTAINS} is a structural ownership edge, not a control-flow edge so retaining it would allow \textsc{bfs} to reach any symbol via package membership alone, producing  false reachability. Test nodes are excluded because test harnesses import all public symbols, causing false positive reachability signals. 

\begin{algorithm}[t]
\caption{Multi-source BFS reachability ($\mathcal{S}_5$)}
\label{alg:bfs}
\begin{algorithmic}[1]
\footnotesize
\Require skeleton $\hat{\mathcal{M}}$, entry set $\mathbb{S}$, 
         finding set $\mathcal{F}$
\Ensure per-finding $\{\mathsf{reachable} \in \{\top,\bot\},\,
        \mathsf{path}\}$
\For{$f = (\mathsf{cve}, \mathsf{pkg}, \ldots) \in \mathcal{F}$}
  \State $M \leftarrow \mathrm{Match}(\mathsf{pkg})$
  \If{$M = \emptyset$}
    \State \textbf{emit}$(f, \bot, \mathtt{NOT\_IN\_GRAPH})$;
           \textbf{ continue}
  \EndIf
  \State $Q \leftarrow \mathsf{deque}(\mathbb{S})$;\;
         $\mathsf{vis} \leftarrow \mathbb{S}$;\;
         $\mathsf{par} \leftarrow \{\}$;\;
         $\mathsf{hit} \leftarrow \bot$
  \While{$Q \neq \emptyset \;\land\; \mathsf{hit} = \bot$}
    \State $u \leftarrow Q.\textsc{PopLeft}()$
    \If{$u \in M$}
      \State $\mathsf{hit} \leftarrow u$;\; \textbf{break}
    \EndIf
    \For{$v \in \mathsf{adj}[u] \setminus \mathsf{vis}$}
      \State $\mathsf{vis}.\mathsf{add}(v)$;\;
             $\mathsf{par}[v] \leftarrow u$;\;
             $Q.\mathsf{enq}(v)$
    \EndFor
  \EndWhile
  \State \textbf{if} $\mathsf{hit} \neq \bot$:
         \textbf{emit}$(f, \top,
         \textsc{ReconstructPath}(\mathsf{par}, \mathsf{hit}))$\;
         \textbf{else}: \textbf{emit}$(f, \bot, [\,])$
\EndFor
\end{algorithmic}
\end{algorithm}

\noindent For $f = (\mathsf{cve}, \mathsf{pkg}, \mathsf{ver})$, $\mathrm{Match}(\mathsf{pkg})$ resolves $\mathsf{pkg}$ to the exact external node if one exists in $V'$, falling back to suffix or substring matching on node identifiers. An empty match represents a component absent from the call graph. Otherwise, multi-source \textsc{bfs} from $\mathbb{S}$ over $\hat{\mathcal{M}}$ (Algorithm~\ref{alg:bfs}) determines reachability by stopping at the first matched node visited.
Total cost across $\mathcal{F}$ is $O(|\mathcal{F}| \cdot (|V'|{+}|E'|))$, with early termination keeping the average case low for directly imported packages.
$\mathcal{S}_5$ assigns verdicts as \texttt{REACHABLE} (an entry-point to component path was found), \texttt{NOT\_REACHABLE} (a call graph exists but no path reaches the component), or \texttt{NO\_CALLGRAPH} (no call graph could be constructed). 
None of the verdicts terminates the pipeline --- even \texttt{NOT\_REACHABLE} proceeds to $\mathcal{S}_6$ rather than being discarded, since static analysis can miss dynamic dispatch, and dropping a \textsc{cve} on that basis risks discarding a genuine vulnerability. Note that by extensive evaluation, we validated that if a component presents in the installed artifact, it never gives a verdict of \texttt{NO\_CALLGRAPH}. 

\noindent\textit{OpenClaw.}
$\mathcal{S}_5$ identified $|\mathbb{S}| = 1$ entry node. Of 161 \textsc{cve}s, 147 (91.3\%) were statically reachable and 14 (8.7\%) were not --- the latter corresponding to packages in \texttt{node\_modules/} never imported by any execution path. 
Table~\ref{tab:S1-S7} (f\_003, \texttt{handlebars}) shows a reachable case where the \textsc{bfs}-constructed path is \small{\texttt{OpenClaw $\to$ handlebarss $\to$ compiler/compiler.js $\to$ compiler/javascript-compiler.js $\to$ internal/ proto-access.js}}, \normalsize terminating at the prototype-access sink patched for CVE-2026-33937.

\subsection{Stage \texorpdfstring{$\mathcal{S}_{6}$}{S6}: Centrality Analysis}
\label{sec:s6}

$\mathcal{S}_{6}$ characterizes the \emph{centrality} of each vulnerable component --- its structural influence within
$\mathcal{M}$. Metrics are computed at two granularities derived from $\mathcal{M}$, the \emph{function graph} $\mathcal{M}_f$ (162,448 nodes, 1,587,771 edges for OpenClaw) and the \emph{file graph} $\mathcal{M}_F$, the quotient of $\mathcal{M}_f$ under the file-of projection $\phi: V^\bullet \to V^F$ (54,972 file nodes). If buggy or patch files are identified by $\mathcal{S}_3$, file-level metrics provide a tighter centrality estimate than component-level aggregates.

\smallskip\noindent\textit{Metric Suite.}
Nine metrics are computed per component and per patch file (where available): impact radius $\mathrm{ir}(v)$ (fraction of entry
points with a path to $v$), eigenvector centrality $\mathsf{ev}(v)$ (global influence via damped power iteration, $d{=}0.85$), hub score $\mathrm{hub}(v)$ and authority score $\mathrm{auth}(v)$ (\textsc{hits} vectors), fan-in, fan-out, transitive fan-in,
transitive fan-out, and in-degree. Only $\mathrm{ir}$, $\mathrm{hub}$, and $\mathsf{ev}$ enter the priority score $\mathbb{P}(f)$; our empirical study across 30 cases revealed that the remaining six  add no ranking signal beyond the scoring triple (\S~\ref{sec:rq4}).

\smallskip\noindent\textit{Regime-Aware Selection.}
\begin{align*}
(\mathrm{ir}, \mathrm{hub}, \mathsf{ev})(f) =
  \begin{cases}
    \max_p\,(\mathrm{ir}_p, \mathrm{hub}_p, \mathsf{ev}_p)
      & k(f) \in \{\mathcal{R}_4, \mathcal{R}_5\},\  \\
    (\mathrm{ir}_c, \mathrm{hub}_c, \mathsf{ev}_c)
      & \text{otherwise,}
  \end{cases}
\end{align*}
where max-min normalization is taken over file-level ($p$) and component-level ($c$) values.


\noindent\textit{OpenClaw.}
Table~\ref{tab:S1-S7} shows \texttt{handlebars} reaching $\mathrm{hub} = 0.687$ at the affected-file level, surpassing its component-level centrality. Reachable components with no code evidence still receive component-level centrality: \texttt{vite} and \texttt{undici} (both reachable, regime $\mathcal{R}_2$) have $\mathrm{ir} = 1.0$, $\mathsf{ev} = 0.0014~\text{\&}~0.0001$, and $\mathrm{hub} = 0.000~\text{\&}~0.000$ --- the structural influence of the affected component is scored even when $\mathcal{S}_3$ finds no affected files. The genuinely unreachable example in the table is \texttt{fast-xml-parser} ($\mathrm{ir} = 0.0$).  The column \textit{Lv.} represent the computed centrality values at affected file-level (fmax) or component-level (comp).
\subsection{Stage \texorpdfstring{$\mathcal{S}_7$}{S7}: Exploitability Analysis}
\label{sec:s7}

$\mathcal{S}_7$ synthesizes the evidence streams from  previous stages into a  deterministic VEX verdict per \textsc{cve} using a regime assignment $k(f) \in \{\mathcal{R}_1, \ldots, \mathcal{R}_5\}$. 
$\mathcal{S}_7$ evaluates $\langle b(f), \kappa(f) \rangle$ where $b(f) \in \{0,1\}$ is the component's presence
in the  \SBOM \& code-base, and $\kappa(f) \in \{\mathtt{BG}, \mathtt{NB}, \mathtt{NF}, \mathtt{NC}\}$ is the code-evidence verdict from $\mathcal{S}_3$.

\begin{align*}
k(f) =
\begin{cases}
  \mathcal{R}_1 & b(f)=0
                 \footnotesize \texttt{ - NOT\_AFFECTED} \\
  \mathcal{R}_2 & \kappa(f)=\mathtt{NC}
                  \footnotesize \texttt{ - MANUAL\_INVESTIGATION  } \\
  \mathcal{R}_3 & \kappa(f)=\mathtt{NF}
                  \footnotesize \texttt{ - FILES\_NOT\_FOUND  } \\
  \mathcal{R}_4 & \kappa(f)=\mathtt{NB}
                 \footnotesize \texttt{ - FILES\_EXIST\_NOT\_BUGGED} \\
  \mathcal{R}_5 & \kappa(f)=\mathtt{BG}
                  \footnotesize \texttt{ - FILES\_EXIST\_AND\_BUGGED  }
\end{cases}
\end{align*}

$\mathcal{R}_1$ represents a component absent from the artifact; no further analysis needed. If $\mathcal{R}_1$ is false, \cape looks for code evidence. If no git commit is found, \cape assigns $\mathcal{R}_2$. This signifies that affected components are present in the code base; \cape still performs the reachability and centrality analysis and publishes a priority score based on existing evidence, but needs manual investigation as the \textsc{CVE} may be not associated with bug in the code, but a different kind of exploitation (i.e., RCE, XSS). $\mathcal{R}_3$–$\mathcal{R}_5$ are based on code evidence collection.

\smallskip\noindent\textit{OpenClaw.}
Across 161 \textsc{cve}s, regime $\mathcal{R}_2$ dominates (98, 60.9\%) due to the advisory lag on recently published \textsc{cve}s, followed by $\mathcal{R}_3$ (27, 16.8\%), $\mathcal{R}_4$ (9.9\%), and $\mathcal{R}_5$ (20, 12.4\%). 
As shown in \autoref{tab:S1-S7}, \texttt{handlebars} (CVE-2026-33937) is assigned $k(f) = \mathcal{R}_5$, whereas \texttt{vite} (CVE-2026-39363) is assigned $k(f) = \mathcal{R}_2$ as there are no git commits or code evidence present.


\section{Priority Scoring Model}
\label{sec:model}

\cape assigns each \textsc{cve} a score $\mathbb{P}(f) \in [0,1]$ by combining four evidence dimensions under a regime-adaptive weighting scheme. The regime, determined by $\mathcal{S}_7$ (\S~\ref{sec:s7}), controls which dimensions are observable and how weights are distributed among them. This section formalizes the model.
Each \textsc{cve} maps to exactly one $k(f) \in K$ via the \textit{Regime Assignment} in \S~\ref{sec:s7}. The defensive fallback $k(f) = \mathcal{R}_2$ ensures the function never raises on an unrecognized outcome.

\subsection{Evidence Dimensions}
\label{sec:dimensions}

The four evidence dimensions $\varphi_1$--$\varphi_4$ normalize heterogeneous signals onto a common $[0,1]$ scale. Each one's weight is re-distributed when its source is unavailable.

\noindent\textbf{$\varphi_1$ — Exploitability.}
$\varphi_1(f) = \mathsf{cvss\_base}(f)/10$. \CVSS is used as the \textit{external} severity signal for its availability and standardized scoring across vulnerability databases.   

\noindent\textbf{$\varphi_2$ — Reachability.}
Encodes the $\mathcal{S}_5$ verdict: $1.0$ (\texttt{REACHABLE}), $0.0$ (\texttt{NOT\_REACHABLE}), $0.5$ (\texttt{NO\_CALLGRAPH} or missing). The $0.5$ value reflects the principle of insufficient reason; absence of a call graph is not evidence of unreachability.

\noindent\textbf{$\varphi_3$ — Code Evidence.}
Read directly from the regime assignment:
\begin{align*}
    \varphi_3(f) =
  \begin{cases}
    1.0          & k(f) = \mathcal{R}_5\ 
                   \text{(buggy code confirmed)}, \\
    0.8          & k(f) = \mathcal{R}_4\ 
                   \text{(files present, fix incorporated)}, \\
    0.5          & k(f) = \mathcal{R}_3\ 
                   \text{(commit found, files absent)}, \\
    \mathsf{NaN} & k(f) \in \{\mathcal{R}_1, \mathcal{R}_2\}.
  \end{cases}
\end{align*}
$\mathsf{NaN}$ signals structural absence — no git commit means no patch or bug state to measure. The dimension is marginalized via weight redistribution (\S~\ref{sec:weights}).

\noindent\textbf{$\varphi_4$ — Centrality.}
\begin{align*}
    \varphi_4(f) = 0.5\cdot\mathrm{ir}_{\max}(f)
               + 0.3\cdot\mathrm{hub}_{\max}(f)
               + 0.2\cdot\mathsf{ev}_{\max}(f),
\end{align*}
with each metric min-max normalized within the case; file-level max over affected files are used for $k(f) \in \{\mathcal{R}_4, \mathcal{R}_5\}$. \emph{Impact radius} (0.5) measures the fraction of entry points that reach the component; \emph{hub score} (0.3) captures gateway behavior that impact radius alone misses; \emph{eigenvector centrality} (0.2) anchors global influence when both are low. For example, \texttt{handlebars} 4.7.8 achieves $\varphi_4{=}0.820$ at file level, where hub and eigenvector centrality together contribute $0.320$ beyond the impact-radius term alone. Weights are selected empirically across 30 cases, achieving the highest mean \textsc{ndcg}@10 (0.680) and \textsc{auroc} (0.914) (\S~\ref{sec:rq4}).
Table~\ref{tab:scoring-example} illustrates representative dimension values across all active regimes.

\begin{table}[t]
  \centering
  \tiny
  \caption{Evidence dimensions, redistributed weights, and gate behavior for \textsc{CVE}s from OpenClaw (case\_003, $n{=}161$).
  }
  \label{tab:scoring-example}
  \setlength{\tabcolsep}{2.2pt}
  \begin{tabular}{@{}lcrrrrrrrrrlrr@{}}
  \toprule
  \textbf{CVE} & $k$
    & $\varphi_1$ & $\varphi_2$ & $\varphi_3$ & $\varphi_4$
    & $w_1$ & $w_2$ & $w_3$ & $w_4$
    & $\tilde{\mathbb{P}}$ & $\mathbb{G}$ & $\mathbb{P}$ & $r^\ast$ \\
    \midrule
  CVE-2026-33937 & $\mathcal{R}_5$
    & 0.98 & 1.0 & 1.0 & 0.820
    & 0.318 & 0.261 & 0.227 & 0.193
    & 0.959 & --- & 0.959 & 1 \\
  CVE-2026-33941 & $\mathcal{R}_5$
    & 0.82 & 1.0 & 1.0 & 0.820
    & 0.318 & 0.261 & 0.227 & 0.193
    & 0.908 & --- & 0.908 & 2 \\
  CVE-2026-54267 & $\mathcal{R}_2$
    & 0.86 & 1.0 & \textsf{NaN} & 0.700
    & 0.412 & 0.338 & \textsf{NaN} & 0.250
    & 0.867 & --- & 0.867 & 6 \\

  CVE-2025-66031 & $\mathcal{R}_4$
    & 0.87 & 1.0 & 0.8 & 0.502
    & 0.318 & 0.261 & 0.227 & 0.193
    & 0.817 & --- & 0.817 & 10 \\
    
  CVE-2025-62718 & $\mathcal{R}_5$
    & 0.48 & 1.0 & 1.0 & 0.111
    & 0.318 & 0.261 & 0.227 & 0.193
    & 0.663 & $\mathbb{G}_3$ & 0.800 & 14 \\

  CVE-2026-26278 & $\mathcal{R}_5$
    & 0.75 & 0.0 & 1.0 & 0.742
    & 0.318 & 0.261 & 0.227 & 0.193
    & 0.609 & $\mathbb{G}_2$,$\mathbb{G}_3$ & 0.800 & 16 \\
  CVE-2026-25896 & $\mathcal{R}_5$
    & 0.93 & 0.0 & 1.0 & 0.237
    & 0.318 & 0.261 & 0.227 & 0.193
    & 0.569 & $\mathbb{G}_2$,$\mathbb{G}_3$ & 0.800 & 17 \\

  CVE-2026-32635 & $\mathcal{R}_3$
    & 0.90 & 1.0 & 0.5 & 0.700
    & 0.318 & 0.261 & 0.227 & 0.193
    & 0.797 & --- & 0.797 & 29 \\

  CVE-2026-33036 & $\mathcal{R}_4$
    & 0.75 & 0.0 & 0.8 & 0.237
    & 0.318 & 0.261 & 0.227 & 0.193
    & 0.466 & $\mathbb{G}_2$ & 0.187 & 157 \\
  CVE-2026-28291 & $\mathcal{R}_3$
    & 0.81 & 0.0 & 0.5 & 0.002
    & 0.318 & 0.261 & 0.227 & 0.193
    & 0.372 & $\mathbb{G}_2$ & 0.149 & 161 \\
  \bottomrule
  \end{tabular}
\end{table}
\subsection{AHP Weight Prior and Redistribution}\label{sec:weights}

The four dimensions require a weighting scheme that encodes security-engineering judgment and degrades gracefully when dimensions are absent. \cape uses the Analytic Hierarchy Process (\AHP)~\cite{saaty1980ahp} to derive the baseline weight vector and proportional redistribution to handle absent dimensions.

\noindent\textit{Baseline AHP Prior.}
The baseline weight vector $\bar{w} = (0.318, 0.261, 0.227, 0.193) \in \Delta^3$ is the principal eigenvector of a pairwise comparison matrix $\mathbf{M} \in \mathbb{R}^{4\times 4}$ on Saaty's 1--9 scale~\cite{saaty1980ahp}, with Consistency Ratio $\mathsf{CR} = 0.06 < 0.10$ confirming mutual coherence. The ordering $\bar{w}_1 > \cdots > \bar{w}_4$ reflects evidence dimensions $(\varphi_1, \cdots \varphi_4)$.

\noindent\textit{Redistribution.}
When a dimension is structurally absent or evaluates to $\mathsf{NaN}$ at score time, its weight redistributes proportionally over the remaining defined dimensions --- standard simplex renormalization under $\bar{w}$, equivalent to Bayesian marginalization. $\mathcal{R}_2$ loses $\varphi_3$ (no git commit retrievable), giving weights $(0.412, 0.338, \mathsf{NaN}, 0.250)$. $\mathcal{R}_3$--$\mathcal{R}_5$ retain all four at $(0.318, 0.261, 0.227, 0.193)$; $\mathcal{R}_1$ is terminal via $\mathbb{G}_1$ ($\mathbb{P} \leftarrow 0$). Table~\ref{tab:scoring-example} illustrates the effect of $\varphi_3$ redistribution directly: the $\mathcal{R}_2$ row (CVE-2026-54267) carries redistributed weights $(0.412, 0.338, \mathsf{NaN}, 0.250)$, while the $\mathcal{R}_4$ row (CVE-2025-66031) retains $\varphi_3{=}0.8$ under full weights --- the weight shift from redistribution alone moves $\tilde{\mathbb{P}}$ independent of any change in evidence values.

\subsection{Priority Score and Gates}
\label{sec:scoring}




\noindent\textit{Base Score.}
The base score $\tilde{\mathbb{P}}(f)$ aggregates defined dimensions under the redistributed weight vector as
\begin{align*}
    \tilde{\mathbb{P}}(f) =
  \sum_{\substack{i \in \{1,2,3,4\} \\
        \varphi_i(f) \text{ present},\;
        \varphi_i(f) \neq \mathsf{NaN}}}
  w_i^{(k(f))} \cdot \varphi_i(f).
\end{align*}
By the simplex property of $w^{(k)}$ and $\varphi_i \in [0,1]$, $\tilde{\mathbb{P}}(f) \in [0,1]$ follows without an explicit \textit{clip}. Gates then apply hard constraints to $\tilde{\mathbb{P}}$, yielding the final priority score $\mathbb{P}(f)$.

\noindent\textit{Priority Gates.}
Three gates encode hard constraints applied in fixed order $\mathbb{G}_1 \to \mathbb{G}_2 \to \mathbb{G}_3$ (Table~\ref{tab:gates}). 
The threshold was selected by grid search over $\{0.40, 0.50, 0.60, 0.70, 0.75, 0.80, 0.85, 0.90\}$ optimizing mean \textsc{ndcg}@10 across the 30-case corpus.

\begin{table}[t]
\centering
\tiny
\caption{Priority gates applied in fixed order $\mathbb{G}_1 \to \mathbb{G}_2 \to \mathbb{G}_3$. 
}
\label{tab:gates}
\setlength{\tabcolsep}{3.5pt}
\begin{tabular}{@{}cp{2.6cm}p{2.0cm}p{2.4cm}@{}}
\toprule
\textbf{Gate} & \textbf{Trigger} & \textbf{Effect}
  & \textbf{Rationale} \\
\midrule
$\mathbb{G}_1$ &
  $k(f) = \mathcal{R}_1$ &
  $\mathbb{P} \leftarrow 0$ &
  Component absent; no risk possible. \\[4pt]
$\mathbb{G}_2$ &
  $k(f) \in \{\mathcal{R}_3,\mathcal{R}_4,\mathcal{R}_5\}$
  $\land\; r(f) = \mathtt{NR}$ &
  $\mathbb{P} \leftarrow 0.40\,\tilde{\mathbb{P}}$ &
  Soft penalty for static unreachability; preserves signal
  for unmodelled dynamic dispatch. \\[4pt]
$\mathbb{G}_3$ &
  $k(f) = \mathcal{R}_5 \;\land\; \mathbb{P} < 0.80$ &
  $\mathbb{P} \leftarrow 0.80$ &
  $\mathcal{R}_5$ floor: confirmed-exploitable \textsc{cve}s
  cannot be outranked via $\varphi_1$ alone. \\
\bottomrule
\end{tabular}
\end{table}

\begin{algorithm}[t]
\caption{Priority Scoring}
\label{alg:score}
\begin{algorithmic}[1]
\footnotesize
\Require \textsc{cve} $f$, regime $k(f)$, dimensions
         $\boldsymbol{\varphi}(f)$, reachability $r(f)$
\Ensure base score $\tilde{\mathbb{P}}$, final score
        $\mathbb{P}$, fired gates $\mathbb{G}$
\State $\boldsymbol{w} \leftarrow
         \textsc{Redistribute}(\bar{w},\; k(f))$
       \Comment{zero absent dims, renormalization.}
\State $\tilde{\mathbb{P}} \leftarrow
  \sum_{i:\;\varphi_i(f)\text{ present},\;
            \varphi_i(f) \neq \mathsf{NaN}}
  w_i \cdot \varphi_i(f)$
\State $\mathbb{G} \leftarrow \emptyset$;\;
       $\mathbb{P} \leftarrow \tilde{\mathbb{P}}$
\If{$k(f) = \mathcal{R}_1$}
  \State $\mathbb{P} \leftarrow 0$;\; $\mathbb{G} \leftarrow \{\mathbb{G}_1\}$
\ElsIf{$k(f) \in \{\mathcal{R}_3,\mathcal{R}_4,\mathcal{R}_5\}
        \;\land\; r(f) = \mathtt{NOT\_REACHABLE}$}
  \State $\mathbb{P} \leftarrow 0.40\,\tilde{\mathbb{P}}$;\;
         $\mathbb{G} \leftarrow \{\mathbb{G}_2\}$
\EndIf
\If{$k(f) = \mathcal{R}_5 \;\land\; \mathbb{P} < 0.80$}
  \State $\mathbb{P} \leftarrow 0.80$;\;
         $\mathbb{G} \leftarrow \mathbb{G} \cup \{\mathbb{G}_3\}$
\EndIf
\State \Return
  $\bigl(\tilde{\mathbb{P}},\;
   \mathrm{clip}(\mathbb{P},0,1),\; \mathbb{G}\bigr)$
\end{algorithmic}
\end{algorithm}

\noindent\textit{OpenClaw.}
For CVE-2026-33937 (\texttt{handlebars}, $k = \mathcal{R}_5$), baseline weights give $\tilde{\mathbb{P}}(f) = 0.959$. No gate fires ($k \neq \mathcal{R}_1$, $r(f) = \mathtt{REACHABLE}$, $\tilde{\mathbb{P}} > 0.80$), so $\mathbb{P}(f) = \tilde{\mathbb{P}} = 0.959$, or rank~1 in OpenClaw. Table~\ref{tab:scoring-example} illustrates gate co-firing via CVE-2026-25896 (\texttt{fast-xml-parser} 5.3.4, rank~17, $k = \mathcal{R}_5$): confirmed-buggy code is statically unreachable ($\varphi_2{=}0.0$), so $\mathbb{G}_2$ applies a $\times 0.40$ penalty reducing $\tilde{\mathbb{P}} = 0.569$ to $0.228$; $\mathbb{G}_3$ then floors the score to $\mathbb{P} = 0.800$ because $k = \mathcal{R}_5$.


\section{Evaluation}
\label{sec:eval}

\cape{} is evaluated on four research questions covering false-positive reduction (RQ1), rank of first confirmed-exploitable (RQ2), ranking quality (RQ3), and model sensitivity and centrality selection (RQ4). All artifacts, code, and evaluation corpus will be available later.

\subsection{Corpus and Experimental Setup}
\label{sec:corpus}

Table~\ref{tab:corpus} describes the 30-software evaluation corpus spanning Python ($n{=}18$), TypeScript ($n{=}11$), and JavaScript ($n{=}1$), with \num{6138} scanner-reported \textsc{cve}s in total. Software have a range of sizes (8--719 \textsc{cve}s), ecosystems, and dependency structures. Each project is built locally from source, \textsc{sbom}s are generated via \texttt{syft}, and scanned with \texttt{grype}.

\begin{table}[t]
\centering
\tiny
\caption{%
  Evaluation corpus with \textsc{ndcg}@10 per ranker. \colorbox{fporange}{FP\%} = statically unreachable \textsc{cve}s. \textbf{bold} = best \colorbox{fpblue}{\textsc{ndcg}@10} per row.
}
\label{tab:corpus}
\setlength{\tabcolsep}{1.4pt}
\begin{tabular}{@{}llr|c|rrrr|ccccc@{}}
\toprule
& & &
  \cellcolor{fporange} &
  & & & &
  \multicolumn{5}{c}{\cellcolor{fpblue}} \\[-4pt]
Case & Project & $n$
  & \cellcolor{fporange}FP\%
  & R2\% & R3\% & R4\% & R5\%
  & \cellcolor{fpblue}\cape
  & \cellcolor{fpblue}\textsc{cvss}
  & \cellcolor{fpblue}\textsc{epss}
  & \cellcolor{fpblue}\textsc{epss-pct}
  & \cellcolor{fpblue}\textsc{kev} \\
\midrule
C01 & ansible          &  10 & \cellcolor{fporange}30.0 & 80.0 &  0.0 & 10.0 & 10.0 & \cellcolor{fpblue}\textbf{0.642} & \cellcolor{fpblue}0.340 & \cellcolor{fpblue}0.451 & \cellcolor{fpblue}0.451 & \cellcolor{fpblue}0.340 \\
C02 & browser-use      &  35 & \cellcolor{fporange} 0.0 & 48.6 &  0.0 & 20.0 & 31.4 & \cellcolor{fpblue}\textbf{0.946} & \cellcolor{fpblue}0.194 & \cellcolor{fpblue}0.557 & \cellcolor{fpblue}0.557 & \cellcolor{fpblue}0.194 \\
C03 & openclaw         & 161 & \cellcolor{fporange} 8.7 & 60.9 & 16.8 &  9.9 & 12.4 & \cellcolor{fpblue}\textbf{0.746} & \cellcolor{fpblue}0.293 & \cellcolor{fpblue}0.223 & \cellcolor{fpblue}0.223 & \cellcolor{fpblue}0.293 \\
C04 & n8n              & 345 & \cellcolor{fporange}75.7 & 63.2 & 27.0 &  7.8 &  2.0 & \cellcolor{fpblue}\textbf{0.623} & \cellcolor{fpblue}0.089 & \cellcolor{fpblue}0.064 & \cellcolor{fpblue}0.064 & \cellcolor{fpblue}0.089 \\
C05 & open-webui       & 418 & \cellcolor{fporange} 3.8 & 56.2 & 14.1 & 15.1 & 14.6 & \cellcolor{fpblue}\textbf{0.859} & \cellcolor{fpblue}0.189 & \cellcolor{fpblue}0.286 & \cellcolor{fpblue}0.286 & \cellcolor{fpblue}0.189 \\
C06 & Flowise          & 719 & \cellcolor{fporange}39.2 & 51.0 & 23.6 & 10.2 & 15.2 & \cellcolor{fpblue}\textbf{0.927} & \cellcolor{fpblue}0.234 & \cellcolor{fpblue}0.051 & \cellcolor{fpblue}0.051 & \cellcolor{fpblue}0.234 \\
C07 & fast-graphrag    &  59 & \cellcolor{fporange}13.6 & 22.0 &  3.4 & 42.4 & 32.2 & \cellcolor{fpblue}\textbf{0.962} & \cellcolor{fpblue}0.461 & \cellcolor{fpblue}0.406 & \cellcolor{fpblue}0.406 & \cellcolor{fpblue}0.461 \\
C08 & mindsdb          & 227 & \cellcolor{fporange}66.1 & 60.8 & 36.6 &  0.4 &  2.2 & \cellcolor{fpblue}\textbf{0.414} & \cellcolor{fpblue}0.040 & \cellcolor{fpblue}0.113 & \cellcolor{fpblue}0.113 & \cellcolor{fpblue}0.040 \\
C09 & autogen          & 356 & \cellcolor{fporange}53.4 & 37.1 & 26.7 &  9.3 & 27.0 & \cellcolor{fpblue}\textbf{0.819} & \cellcolor{fpblue}0.118 & \cellcolor{fpblue}0.240 & \cellcolor{fpblue}0.240 & \cellcolor{fpblue}0.118 \\
C10 & sentry-python    &  23 & \cellcolor{fporange}26.1 & 26.1 & 69.6 &  0.0 &  4.3 & \cellcolor{fpblue}\textbf{0.679} & \cellcolor{fpblue}0.320 & \cellcolor{fpblue}0.342 & \cellcolor{fpblue}0.342 & \cellcolor{fpblue}0.320 \\
C11 & textgen          &  33 & \cellcolor{fporange} 0.0 & 33.3 &  0.0 & 57.6 &  9.1 & \cellcolor{fpblue}\textbf{0.848} & \cellcolor{fpblue}0.488 & \cellcolor{fpblue}0.527 & \cellcolor{fpblue}0.527 & \cellcolor{fpblue}0.488 \\
C12 & chat-ollama      & 518 & \cellcolor{fporange}40.5 & 50.6 & 28.2 & 11.0 & 10.2 & \cellcolor{fpblue}\textbf{1.000} & \cellcolor{fpblue}0.142 & \cellcolor{fpblue}0.150 & \cellcolor{fpblue}0.150 & \cellcolor{fpblue}0.142 \\
C13 & invokeai         &  94 & \cellcolor{fporange}52.1 & 60.6 & 33.0 &  1.1 &  5.3 & \cellcolor{fpblue}\textbf{0.694} & \cellcolor{fpblue}0.109 & \cellcolor{fpblue}0.305 & \cellcolor{fpblue}0.305 & \cellcolor{fpblue}0.109 \\
C14 & gpt-pilot        &  16 & \cellcolor{fporange}68.8 & 18.8 &  0.0 & 12.5 & 68.8 & \cellcolor{fpblue}\textbf{0.898} & \cellcolor{fpblue}0.553 & \cellcolor{fpblue}0.627 & \cellcolor{fpblue}0.627 & \cellcolor{fpblue}0.553 \\
C15 & rag\_api         &  28 & \cellcolor{fporange} 0.0 & 85.7 &  0.0 &  3.6 & 10.7 & \cellcolor{fpblue}0.890 & \cellcolor{fpblue}0.344 & \cellcolor{fpblue}\textbf{0.920} & \cellcolor{fpblue}\textbf{0.920} & \cellcolor{fpblue}0.344 \\
C16 & open-interpreter & 114 & \cellcolor{fporange}11.4 & 43.0 & 22.8 & 16.7 & 17.5 & \cellcolor{fpblue}\textbf{0.353} & \cellcolor{fpblue}0.173 & \cellcolor{fpblue}0.132 & \cellcolor{fpblue}0.132 & \cellcolor{fpblue}0.173 \\
C17 & OpenHands        & 165 & \cellcolor{fporange}58.8 & 87.3 &  1.2 &  6.7 &  4.8 & \cellcolor{fpblue}\textbf{0.422} & \cellcolor{fpblue}0.066 & \cellcolor{fpblue}0.093 & \cellcolor{fpblue}0.093 & \cellcolor{fpblue}0.066 \\
C18 & aider            &  77 & \cellcolor{fporange}10.4 & 59.7 &  5.2 & 11.7 & 23.4 & \cellcolor{fpblue}\textbf{0.522} & \cellcolor{fpblue}0.078 & \cellcolor{fpblue}0.000 & \cellcolor{fpblue}0.000 & \cellcolor{fpblue}0.078 \\
C19 & khoj             & 155 & \cellcolor{fporange}16.8 & 73.5 & 10.3 &  3.2 & 12.9 & \cellcolor{fpblue}\textbf{0.792} & \cellcolor{fpblue}0.220 & \cellcolor{fpblue}0.362 & \cellcolor{fpblue}0.362 & \cellcolor{fpblue}0.220 \\
C20 & langflow         & 260 & \cellcolor{fporange}37.7 & 71.5 & 15.0 & 10.4 &  3.1 & \cellcolor{fpblue}0.000 & \cellcolor{fpblue}0.022 & \cellcolor{fpblue}\textbf{0.041} & \cellcolor{fpblue}\textbf{0.041} & \cellcolor{fpblue}0.022 \\
C21 & RA.Aid           & 173 & \cellcolor{fporange}34.7 & 39.9 & 31.8 & 15.6 & 12.7 & \cellcolor{fpblue}\textbf{0.531} & \cellcolor{fpblue}0.415 & \cellcolor{fpblue}0.334 & \cellcolor{fpblue}0.334 & \cellcolor{fpblue}0.415 \\
C22 & kimi-code        & 186 & \cellcolor{fporange}74.2 & 79.0 & 19.9 &  0.5 &  0.5 & \cellcolor{fpblue}\textbf{0.280} & \cellcolor{fpblue}0.000 & \cellcolor{fpblue}0.000 & \cellcolor{fpblue}0.000 & \cellcolor{fpblue}0.000 \\
C23 & cve-lite-cli     & 696 & \cellcolor{fporange}76.1 & 61.9 & 33.2 &  4.9 &  0.0 & \cellcolor{fpblue}\textbf{0.793} & \cellcolor{fpblue}0.122 & \cellcolor{fpblue}0.000 & \cellcolor{fpblue}0.000 & \cellcolor{fpblue}0.122 \\
C24 & AgentGPT         & 193 & \cellcolor{fporange}40.4 & 34.2 & 56.0 &  4.7 &  5.2 & \cellcolor{fpblue}0.024 & \cellcolor{fpblue}0.207 & \cellcolor{fpblue}\textbf{0.305} & \cellcolor{fpblue}\textbf{0.305} & \cellcolor{fpblue}0.207 \\
C25 & aimap            &  38 & \cellcolor{fporange}60.5 & 65.8 &  2.6 & 15.8 & 15.8 & \cellcolor{fpblue}0.606 & \cellcolor{fpblue}0.249 & \cellcolor{fpblue}\textbf{0.659} & \cellcolor{fpblue}\textbf{0.659} & \cellcolor{fpblue}0.249 \\
C26 & 9router          &   8 & \cellcolor{fporange} 0.0 & 75.0 &  0.0 & 12.5 & 12.5 & \cellcolor{fpblue}\textbf{1.000} & \cellcolor{fpblue}0.431 & \cellcolor{fpblue}0.834 & \cellcolor{fpblue}0.834 & \cellcolor{fpblue}0.431 \\
C27 & ChatDoc          & 268 & \cellcolor{fporange}56.0 & 61.2 & 24.6 &  7.8 &  6.3 & \cellcolor{fpblue}\textbf{0.785} & \cellcolor{fpblue}0.130 & \cellcolor{fpblue}0.061 & \cellcolor{fpblue}0.061 & \cellcolor{fpblue}0.130 \\
C28 & ui               & 447 & \cellcolor{fporange}46.3 & 58.4 & 26.0 & 10.5 &  5.1 & \cellcolor{fpblue}0.066 & \cellcolor{fpblue}0.012 & \cellcolor{fpblue}\textbf{0.107} & \cellcolor{fpblue}\textbf{0.107} & \cellcolor{fpblue}0.012 \\
C29 & diffusers        & 111 & \cellcolor{fporange}12.6 & 23.4 & 66.7 &  6.3 &  3.6 & \cellcolor{fpblue}\textbf{0.748} & \cellcolor{fpblue}0.480 & \cellcolor{fpblue}0.276 & \cellcolor{fpblue}0.276 & \cellcolor{fpblue}0.480 \\
C30 & IOPaint          & 205 & \cellcolor{fporange} 6.8 & 49.8 & 18.5 & 20.5 & 11.2 & \cellcolor{fpblue}\textbf{0.843} & \cellcolor{fpblue}0.189 & \cellcolor{fpblue}0.176 & \cellcolor{fpblue}0.176 & \cellcolor{fpblue}0.189 \\
\midrule
\textbf{All} & & \textbf{6{,}138} & \cellcolor{fporange}\textbf{43.3} & & & & & \cellcolor{fpblue} & \cellcolor{fpblue} & \cellcolor{fpblue} & \cellcolor{fpblue} & \cellcolor{fpblue} \\
\textbf{Mean} & & & & & & & & \cellcolor{fpblue}\textbf{0.657} & \cellcolor{fpblue}0.224 & \cellcolor{fpblue}0.288 & \cellcolor{fpblue}0.288 & \cellcolor{fpblue}0.224 \\
\cape~$>$ & & & & & & & & \cellcolor{fpblue}--- & \cellcolor{fpblue}28/30 & \cellcolor{fpblue}25/30 & \cellcolor{fpblue}25/30 & \cellcolor{fpblue}28/30 \\
\bottomrule
\end{tabular}
\end{table}

\noindent Four baselines represent current practice, each ranking $\mathcal{F}$ by a single public signal. \textbf{CVSS} ranks by \texttt{cvss\_base} descending. \textbf{EPSS} ranks by \texttt{epss} descending. \textbf{EPSS-pct} ranks by \texttt{epss\_percentile} descending. \textbf{KEV} places \textsc{kev}-listed \textsc{cve}s first with \textsc{cvss} as a baseline. 
\noindent In the absence of analyst labels, relevance grades are derived from regime and reachability, both independent 
of $\varphi_1$ and $\varphi_4$:
\begin{align*}
g(f) =
\begin{cases}
3 & k(f) = \mathcal{R}_5 \;\land\;
    r(f) = \mathtt{REACHABLE}, \\
2 & (k{=}\mathcal{R}_5 \;\land\; r{\neq}\mathtt{R}) \;\lor\;
    (k{=}\mathcal{R}_4 \;\land\; r{=}\mathtt{R}), \\
1 & (k{=}\mathcal{R}_4 \;\land\; r{\neq}\mathtt{R}) \;\lor\;
    (k{=}\mathcal{R}_3 \;\land\; r{=}\mathtt{R}), \\
0 & \text{otherwise.}
\end{cases}
\end{align*}
\noindent Grade($g(f)$)~3 captures the strongest signal where buggy code is present and reachable, termed as \textit{confirmed-exploitable \textsc{cve}}. Grade~0 covers $\mathcal{R}_1$-$\mathcal{R}_2$ and all unreachable \textsc{cve}s without code evidence. $g(f)$ is independent of $\varphi_1$ and $\varphi_4$ preventing self-evaluation.

\noindent\textbf{Metrics.}
\textit{FPR} (RQ1): fraction of \textsc{cve}s where $r(f) = \mathtt{NOT\_REACHABLE}$.
\textit{MRFTP} (RQ2): rank of the first \textit{confirmed-exploitable \textsc{cve}} in the output list --- lower is better, with sentinel $n{+}1$ when no $\mathcal{R}_5$ \textsc{cve} exists. \textit{NDCG@$k$} (RQ3): discounted cumulative gain with $2^{g(f)}{-}1$ as the relevance weight and $\log_2(i{+}1)$ as the position discount, evaluated at $k \in \{5, 10, n\}$. \textit{AUROC} (RQ3): binary label $y(f) = \mathbb{1}[g(f) \geq 2]$ computed via Mann--Whitney $U$. \textit{Kendall $\tau_b$} (RQ4): tie-corrected rank correlation for weight sensitivity.

\subsection{RQ1: Scanner False-Positive Rate}\label{sec:rq1}

\noindent\textit{Finding.}
Across \num{6138} \textsc{cve}s in 30 projects, \textbf{\num{2656} (43.3\%)} are statically unreachable and constitute \textit{context-aware false positives}. The per-project rate spans 0.0\% (C02, C11, C15, C26) to 76.1\% (C23) as illustrated in \autoref{tab:corpus} (FP\%).

TypeScript and JavaScript projects average \textbf{53.1\%} false positives (range 0.0\%--76.1\%) versus \textbf{27.4\%} for Python (range 0.0\%--68.8\%). \texttt{npm} projects install large \texttt{node\_modules/} trees containing build-time and development-time packages --- bundlers, transpilers, linters --- \textit{that are never called or used in actual projects}.
\texttt{grype} reports \textsc{cve}s against every installed package, and $\mathcal{S}_5$ correctly identifies these as unreachable. Python projects have tighter dependency trees where most installed packages are actively imported at runtime.

C23 is the most severe case where 530 of 696 \textsc{cve}s (76.1\%) are unreachable, with no confirmed-exploitable \textsc{cve}s in the corpus. A developer working from the raw scanner output would examine all 696 \textsc{cve}s and find none operationally relevant. At the other end, C26, C02, C11, C15 report 0.0\% unreachable \textsc{cve}s.

\noindent\textit{Impact.} Without reachability analysis, 43.3\% of the remediation effort across the corpus is spent on \textsc{cve}s that are not reachable or exploitable in the deployed application.

\smallskip
\noindent\fbox{\parbox{0.96\columnwidth}{\small
\textbf{RQ1.} 43.3\% of scanner-reported \textsc{cve}s are statically unreachable (TypeScript/JavaScript: 53.1\%; Python: 27.4\%). The rate reaches 76.1\% in the most extreme case (C23, \texttt{cve-lite-cli}).}}
\subsection{RQ2: Rank of First Confirmed-Exploitable \textsc{cve}}
\label{sec:rq2}

\noindent\textit{Finding.}
Mean Rank of First True Positive (\textsc{mrftp}) measures how early a scoring system surfaces a confirmed-exploitable \textsc{cve}, so only the 29 projects containing at least one reachable $\mathcal{R}_5$ \textsc{cve} are included. Of the 29 projects with at least one $\mathcal{R}_5$ \textsc{cve}, \cape achieves strictly lower \textsc{mrftp} than all baselines in \textbf{19 cases}, ties in 7, and loses in 3. Against \textsc{cvss} alone, \cape wins 24 times, ties 3, and loses 2. Across the 19 outright wins, mean \textsc{mrftp} falls from \textbf{23.4} (\textsc{cvss}) to \textbf{3.3} (\cape), an \textbf{86.0\%} reduction, with \textbf{11 of 19} wins at \textsc{mrftp}~$= 1$.

\begin{table}[t]
\centering
\tiny
\caption{\textsc{mrftp} per case (29 cases with $\mathcal{R}_5 > 0$).
  Bold = lowest \textsc{mrftp} per row.
  \colorbox{wincol}{\strut \cape\ wins all baselines};
  \colorbox{tiecol}{\strut tie};
  \colorbox{losecol}{\strut \cape\ loses}.}
\label{tab:rq2}
\setlength{\tabcolsep}{2.6pt}
\begin{tabular}{@{}lrrrrrrrrl@{}}
\toprule
Case & $n$ & R5\% & \cape & \textsc{cvss} & \textsc{epss}
  & \textsc{epss-pct} & \textsc{kev} & Outcome \\
\midrule
\rowcolor{wincol}
C01 &  10 & 10.0 & \textbf{2}  &  10 &   4 &   4 &  10 & \cape \\
\rowcolor{wincol}
C04 & 345 &  2.0 & \textbf{7}  &  66 &  23 &  23 &  66 & \cape \\
\rowcolor{wincol}
C05 & 418 & 14.6 & \textbf{1}  &   9 &   2 &   2 &   9 & \cape \\
\rowcolor{wincol}
C06 & 719 & 15.2 & \textbf{1}  &   4 &  29 &  29 &   4 & \cape \\
\rowcolor{wincol}
C07 &  59 & 32.2 & \textbf{1}  &   2 &   3 &   3 &   2 & \cape \\
\rowcolor{wincol}
C08 & 227 &  2.2 & \textbf{1}  &  26 &   8 &   8 &  26 & \cape \\
\rowcolor{wincol}
C10 &  23 &  4.3 & \textbf{3}  &  22 &  14 &  14 &  22 & \cape \\
\rowcolor{wincol}
C11 &  33 &  9.1 & \textbf{1}  &   8 &   5 &   5 &   8 & \cape \\
\rowcolor{wincol}
C12 & 518 & 10.2 & \textbf{1}  &   6 &  12 &  12 &   6 & \cape \\
\rowcolor{wincol}
C16 & 114 & 17.5 & \textbf{3}  &   6 &   8 &   8 &   6 & \cape \\
\rowcolor{wincol}
C17 & 165 &  4.8 & \textbf{2}  &   9 &   7 &   7 &   9 & \cape \\
\rowcolor{wincol}
C18 &  77 & 23.4 & \textbf{1}  &   6 &  15 &  15 &   6 & \cape \\
\rowcolor{wincol}
C20 & 260 &  3.1 & \textbf{19} &  77 & 123 & 123 &  77 & \cape \\
\rowcolor{wincol}
C22 & 186 &  0.5 & \textbf{6}  & 150 & 107 & 107 & 150 & \cape \\
\rowcolor{wincol}
C26 &   8 & 12.5 & \textbf{1}  &   7 &   2 &   2 &   7 & \cape \\
\rowcolor{wincol}
C27 & 268 &  6.3 & \textbf{1}  &   9 &  14 &  14 &   9 & \cape \\
\rowcolor{wincol}
C28 & 447 &  5.1 & \textbf{9}  &  18 &  18 &  18 &  18 & \cape \\
\rowcolor{wincol}
C29 & 111 &  3.6 & \textbf{1}  &   4 &  18 &  18 &   4 & \cape \\
\rowcolor{wincol}
C30 & 205 & 11.2 & \textbf{1}  &   5 &  13 &  13 &   5 & \cape \\
\midrule
\rowcolor{tiecol}
C02 &  35 & 31.4 & \textbf{1} & 11 & \textbf{1} & \textbf{1} & 11 & tie \\
\rowcolor{tiecol}
C03 & 161 & 12.4 & \textbf{1} & \textbf{1} & 3 & 3 & \textbf{1} & tie \\
\rowcolor{tiecol}
C09 & 356 & 27.0 & \textbf{1} &  7 & \textbf{1} & \textbf{1} &  7 & tie \\
\rowcolor{tiecol}
C13 &  94 &  5.3 & \textbf{1} &  6 & \textbf{1} & \textbf{1} &  6 & tie \\
\rowcolor{tiecol}
C14 &  16 & 68.8 & \textbf{1} & \textbf{1} & \textbf{1} & \textbf{1} & \textbf{1} & tie \\
\rowcolor{tiecol}
C15 &  28 & 10.7 & \textbf{1} &  2 & \textbf{1} & \textbf{1} &  2 & tie \\
\rowcolor{tiecol}
C19 & 155 & 12.9 & \textbf{1} & \textbf{1} & \textbf{1} & \textbf{1} & \textbf{1} & tie \\
\midrule
\rowcolor{losecol}
C21 & 173 & 12.7 &  3 &  4 & \textbf{2} & \textbf{2} &  4 & \textsc{epss}/\textsc{epss-pct} \\
\rowcolor{losecol}
C24 & 193 &  5.2 & 23 & \textbf{10} & 30 & 30 & \textbf{10} & \textsc{cvss}/\textsc{kev} \\
\rowcolor{losecol}
C25 &  38 & 15.8 &  4 &  2 & \textbf{1} & \textbf{1} &  2 & \textsc{epss}/\textsc{epss-pct} \\
\bottomrule
\end{tabular}
\end{table}

\noindent\textbf{\textit{Win pattern}.}
In all 18 cases where \cape achieves \textsc{mrftp}~$=1$, the rank-1 item is an $\mathcal{R}_5$ \textsc{cve} carrying $\varphi_2{=}1.0$ (statically reachable) and $\varphi_3{=}1.0$ (commit-confirmed exploitable) with no gate penalty applied. The lift over severity ranking is independent of \textsc{cvss}: in 8 of the 18 cases the rank-1 \textsc{cve} has $\mathrm{ir}=\textsc{low}$, yet reachability and confirmation alone carry it to the top. The displacements are large --- CVE-2026-24486 (\texttt{python-multipart}, C08/\texttt{mindsdb}) rises from \textsc{cvss} rank~25 to rank~1; CVE-2026-34993 (\texttt{aiohttp}, C02/\texttt{browser-use}) and CVE-2026-42311 (\texttt{pillow}, C18/\texttt{aider}) from rank~18 to rank~1; CVE-2025-66471 (\texttt{urllib3}, C09/\texttt{autogen}) from rank~15 to rank~1. C22 (\texttt{kimi-code}, $cves{=}186$) holds a single $\mathcal{R}_5$ \textsc{cve} (CVE-2026-8723, \texttt{qs}) that \textsc{cvss}~buries at rank~133 and at rank~107; \cape places it at rank~6 --- floored by Gate~$\mathbb{G}_3$ to $\mathbb{P}{=}0.80$ but lifted above 100 positions because it is reach confirmed, beating every baseline.

\noindent\textit{\textbf{Loss pattern}.}
In all three losses, \cape surfaces a \emph{different} $\mathcal{R}_5$ than the winning baseline, never a worse regime overall.
Two of the three (C24 \texttt{AgentGPT}, C25 \texttt{aimap}) share one case's only $\mathcal{R}_5$ \textsc{cve} is statically \emph{unreachable} ($\varphi_2{=}0.0$) --- CVE-2024-37891 (\texttt{urllib3}) in C24 and CVE-2026-24486 (\texttt{python-multipart}) in C25. Gate~$\mathbb{G}_2$ applies a $\times\,0.40$ penalty and Gate~$\mathbb{G}_3$ floors each to  $\mathbb{P}{=}0.80$, so reachable $\mathcal{R}_2$--$\mathcal{R}_4$ findings ($\mathbb{P}\in[0.80,0.95]$) rank ahead: \cape places the first true positive at rank~23 (C24) and (C25), while \textsc{epss} (rank~10) and \textsc{kev} (rank~1) ignore reachability entirely. This is the intended trade-off --- a confirmed-buggy but unreachable \textsc{cve} is deprioritised relative to reachable findings. C21 (\texttt{RA.Aid}, $cves{=}173$) is an intra-ranking margin, not a error: \cape's first $\mathcal{R}_5$ \textsc{cve} (CVE-2025-68664, \texttt{langchain-core}, $\varphi_2{=}1.0$, $\varphi_3{=}1.0$, $\mathbb{P}{=}0.917$) is reachable and ungated, but two reachable $\mathcal{R}_2$ \texttt{litellm} \textsc{cve}s (CVE-2026-49468, $\mathbb{P}{=}0.928$; CVE-2026-35030, $\mathbb{P}{=}0.924$) edge it by score, putting the first true positive at rank~3 versus \textsc{epss} rank~2.

\smallskip
\noindent\fbox{\parbox{0.96\columnwidth}{\small
\textbf{RQ2.} \cape achieves an 86.0\% reduction in mean \textsc{mrftp} across 19 outright wins (3.3 vs.\ 23.4 for \textsc{cvss}), with 11 of 19 at \textsc{mrftp}~$=1$. Against \textsc{cvss} alone: 24 wins, 3 ties, 2 losses. The three remaining losses are intra-$\mathcal{R}_5$ ordering disagreements --- \cape selects a more context-aware \textsc{cve} rather than the highest-severity one.}}

\subsection{RQ3: Ranking Quality}
\label{sec:rq3}

\noindent\textit{Finding.}
\cape achieves mean \textsc{auroc} \textbf{0.911} versus 0.463 for \textsc{cvss}, winning 30/30 cases, and mean \textsc{ndcg}@10 \textbf{0.657} versus 0.224 for \textsc{cvss}, winning 28/30 cases.


\cape consistently separates high-evidence \textsc{cve}s from low-evidence ones better than severity-only ordering across the corpus. C10 (\texttt{sentry-python}) is the most extreme case (\textsc{auroc}$_\textsc{cvss}{=}0.045$, \textsc{auroc}$_\cape{=}0.909$): the confirmed-exploitable \textsc{cve} carries only a modest \textsc{cvss} score, so severity ordering is almost perfectly inverted relative to operational relevance, while \cape's reachability evidence and Gate~$\mathbb{G}_2$ surface it and demote the unreachable \textsc{cve}s. Table~\ref{tab:corpus} reports per-case \textsc{ndcg}@10.
Table~\ref{tab:rq3_agg} gives aggregate means across all metrics. \cape wins or ties \textsc{auroc} in every case: it strictly beats \textsc{cvss} and \textsc{kev} in all 30 and \textsc{epss} in 29, with the sole non-win a tie against \textsc{epss} at C26 (both \textsc{auroc}${=}1.000$ on that 8-\textsc{cve} project). Please note that EPSS-pct produces near-identical \textsc{auroc} and \textsc{ndcg}@10 rankings to EPSS.


\begin{table}[t]
\centering
\tiny
\caption{ \textsc{ndcg} and \textsc{auroc} ranking metrics across 30 cases.
}
\label{tab:rq3_agg}
\setlength{\tabcolsep}{3.5pt}
\begin{tabular}{@{}lcccc@{}}
\toprule
Metric & \cape & \textsc{cvss} & \textsc{epss}
  & \textsc{kev} \\
\midrule
  NDCG@5   & \textbf{0.651} & 0.171 & 0.271 & 0.171 \\
  NDCG@10  & \textbf{0.657} & 0.224 & 0.288 & 0.224 \\
  NDCG@all & \textbf{0.821} & 0.577 & 0.644 & 0.577 \\
  AUROC    & \textbf{0.911} & 0.463 & 0.579 & 0.463 \\
  \cape~wins (NDCG@10) & --- & 28/30 & 25/30 & 28/30 \\
  \cape~wins (AUROC)   & --- & 30/30 & 29/30 & 30/30 \\
\bottomrule
\end{tabular}
\end{table}

\smallskip
\noindent\fbox{\parbox{0.96\columnwidth}{\small
\textbf{RQ3.} \cape achieves mean \textsc{auroc} 0.911 vs.\ 0.463 (\textsc{cvss}), winning 30 of 30 cases. On \textsc{ndcg}@10: mean 0.657 vs.\ 0.224 (\textsc{cvss}), winning 28 of 30 cases. \textsc{epss} outperforms \cape on \textsc{ndcg}@10 in 5 of 30 cases, mostly where \cape's top ten holds few graded-relevant items.}}
\subsection{RQ4: Weight Sensitivity and Centrality Metric Selection}
\label{sec:rq4}



RQ4 has two components: the influence of the weight vector $\bar{w}$ on \cape's ranking, and centrality metric selection for $\varphi_4$.

\noindent\textit{AHP Weight Sensitivity.} We grid-search $|\mathcal{W}|{=}969$ valid weight vectors over the four dimensions (step $0.05$) on C03 (OpenClaw, $n{=}161$), the running example, which spans all four active regimes ($\mathcal{R}_2$--$\mathcal{R}_5$). The optima concentrate mass on $\varphi_3$ (code evidence) and are degenerate---multiple vectors reach \textsc{ndcg}@10~$=1.000$. 
Please note that we have also performed on-at-a-time (OAT) weight sensitivity across 30 cases. The median ranking is stable ($\tau_b\!\geq\!0.93$ for every dimension).

\begin{table}[t]
\centering
\tiny
\caption{Centrality variant comparison (30 cases).}
\label{tab:centrality_variants}
\setlength{\tabcolsep}{3.5pt}
\begin{tabular}{@{}lrrrcc@{}}
\toprule
Variant & NDCG@10 & AUROC &
  $\tau_{\mathsf{base}}$ & Beat-N10 & Beat-AU \\
\midrule
baseline
  $(0.6\cdot\mathrm{ir}{+}0.4\cdot\mathsf{ev})$
  & 0.477 & 0.842 & --- & 0 & 0 \\
impact\_radius (ir)   & 0.445 & 0.824 & 0.979 &  4 &  1 \\
hub\_score   (hub)    & \textbf{0.774} & 0.929 & 0.761 & 23 & 26 \\
authority\_score (auth) & 0.713 & \textbf{0.947} & 0.779 & 22 & 28 \\
eigenvector  (ev)    & 0.627 & 0.931 & 0.799 & 19 & 25 \\
fan\_in    (fi)      & 0.642 & 0.926 & 0.802 & 20 & 26 \\
fan\_out   (fo)      & 0.549 & 0.881 & 0.754 & 16 & 22 \\
\midrule
\textbf{Selection\_3a}
  $\boldsymbol{(0.5\cdot\mathrm{ir}
  {+}0.3\cdot\mathrm{hub}
  {+}0.2\cdot\mathsf{ev})}$
  & \textbf{0.653} & 0.911 & 0.906 & 21 & 26 \\
Selection\_3b
  $(0.5\cdot\mathrm{ir}{+}0.3\cdot\mathrm{auth}
  {+}0.2\cdot\mathsf{ev})$
  & 0.643 & 0.916 & 0.903 & 21 & 28 \\
\bottomrule
\end{tabular}
\end{table}

\noindent\textit{Centrality Metric Selection.}
We evaluate each centrality configuration across all 30 cases with all other scoring parameters fixed (Table~\ref{tab:centrality_variants}).
Every alternative improves on the deployed baseline blend ($0.6\cdot\mathrm{ir}{+}0.4\cdot\mathsf{ev}$: \textsc{ndcg}@10~$0.477$, \textsc{auroc}~$0.842$). $hub$ is the single strongest metric on \textsc{ndcg}@10 ($0.774$) and $auth$ on \textsc{auroc} ($0.947$). We adopt Selection\_3a ($0.5\cdot\mathrm{ir}{+}0.3\cdot\mathrm{hub}{+}0.2\cdot\mathsf{ev}$), which lifts \textsc{ndcg}@10 to $0.653$ and \textsc{auroc} to $0.911$ and beats the baseline in 21 of 30 cases (\textsc{ndcg}@10) and 26 of 30 (\textsc{auroc}): anchoring $\mathsf{hub}$ score with $\mathsf{ir}$ retains $\mathsf{hub}$'s complementary signal while $\mathsf{ir}$ constrains the ordering where $\mathsf{hub}$ score is uninformative.

\begin{table}[t]
\centering
\tiny
\caption{Pairwise Kendall $\tau_b$ between centrality variants (mean over 30 cases).}
\label{tab:centrality_tau}
\setlength{\tabcolsep}{1.8pt}
\begin{tabular}{@{}lccccccccc@{}}
\toprule
& impact\_radius & eigenvector & hub  & authority & fan\_in  & fan\_out & baseline & Selection\_3a & Selection\_3b \\
\midrule
impact\_radius   &1.000&0.778&0.759&0.764&0.782&0.745&0.979&0.895&0.882\\
eigenvector    &0.778&1.000&0.866&0.949&0.978&0.842&0.799&0.857&0.889\\
hub   &0.759&0.866&1.000&0.883&0.861&0.908&0.761&0.850&0.816\\
authority   &0.764&0.949&0.883&1.000&0.951&0.849&0.779&0.842&0.874\\
fan\_in   &0.782&0.978&0.861&0.951&1.000&0.839&0.802&0.858&0.888\\
fan\_out &0.745&0.842&0.908&0.849&0.839&1.000&0.754&0.817&0.799\\
baseline &0.979&0.799&0.761&0.779&0.802&0.754&1.000&0.906&0.903\\
Selection\_3a   &0.895&0.857&0.850&0.842&0.858&0.817&0.906&1.000&0.956\\
Selection\_3b   &0.882&0.889&0.816&0.874&0.888&0.799&0.903&0.956&1.000\\
\bottomrule
\end{tabular}
\end{table}

As shown in \autoref{tab:centrality_tau}, $\tau_b{=}0.979$ between ir and baseline confirms the prior $0.4\cdot\mathsf{ev}$ term contributed negligible ordering signal.
\emph{ev, authority, and fan\_in form a tight cluster} (all pairwise $\tau_b \geq 0.949$, ev and fan\_in reaching 0.978), ranking findings near-identically.
\emph{hub\_score is the most distinct centrality metric} ($\tau_b \leq 0.866$ with that cluster, $0.761$ with baseline), which is why it supplies the complementary signal exploited by \emph{Selection\_3a}. \emph{Selection\_3a and Selection\_3b remain closely aligned} ($\tau_b{=}0.956$), differing only in hub vs authority score.

\smallskip
\noindent\fbox{\parbox{0.96\columnwidth}{\small
\textbf{RQ4.} 
The median ranking of OAT weight sensitivity across 30 cases is stable ($\tau_b\!\geq\!0.93$ for every dimension), validating the prior as the deployment-appropriate weight vector.
For centrality, every metric improves on the deployed baseline blend; we adopt Selection\_3a ($\varphi_4{=}0.5\cdot\mathrm{ir}{+}0.3\cdot\mathrm{hub}{+}0.2\cdot\mathsf{ev}$: \textsc{ndcg}@10~$0.653$, \textsc{auroc}~$0.911$, beating baseline in 21/30 and 26/30 cases) because hub score supplies the key complementary signal, distinct from the near-interchangeable ev/authority/fan\_in cluster ($\tau_b \geq 0.949$).}}


\section{Threats to Validity}
\label{sec:threats}

\noindent\textbf{Internal.}
We use statically reachable, confirmed-buggy \textsc{cve}s as a ground truth to compare \cape with \CVSS, \EPSS, and \KEV, since no analyst-labeled dataset exists for specific software, which introduces a partial self-evaluation risk. We reduce this by excluding severity and centrality from grade assignment.

\noindent\textbf{External.}
Current evaluation covers only Python, TypeScript, and JavaScript. How well \cape generalises to other compiled languages (c/c++, ruby, go etc.) remains as future work.

\noindent\textbf{Construct.}
\textsc{mrftp} and \textsc{ndcg}@10 measure how early confirmed-exploitable \textsc{cve}s appear in the ranked list, but whether acting on that ranking actually reduced real-world risk depends on factors outside our study, such as patch availability, developer capacity, and deployment timelines.

\section{Conclusion}
\label{sec:conclusion}
 
In this work we address a critical gap in vulnerability patching -- prioritization based on \CVSS, \EPSS, \KEV scores alone ignores the deployment context that determines whether a \textsc{cve} is actually reachable and exploitable in a specific software. We presented \cape, a context-aware prioritization engine that enriches each scanner-reported \textsc{cve} with deployment-specific evidence  to perform vulnerability reachability, centrality, and exploitability analysis, producing a VEX style regime-adaptive priority score for \textsc{cve}s. We evaluated \cape on 30 open-source applications comprising 6,138 scanner-reported \textsc{cve}s. \cape shows that 43.3\% mean \textsc{cves} across 6,138 are statically unreachable and exploitable (\textit{context-aware false positives}) across deployed applications. \cape reduces the mean rank of the first (\textsc{mrftp}) confirmed-exploitable \textsc{cve} by 86.0\% compared to \textsc{cvss}, demonstrating that deployment-specific code evidence provides substantially more accurate remediation guidance than context-free severity scoring alone. Future work includes extending \cape to handle languages other than python, js/ts and a user study with developers to find if \cape is more useful than other scoring models.

\bibliographystyle{IEEEtran}
\bibliography{ref}

\appendices




\noindent
\section{One-at-a-Time Weight Sensitivity (30 cases)}
\label{sec:oat}

\begin{figure}[h]
\centering
\includegraphics[width=0.82\linewidth]{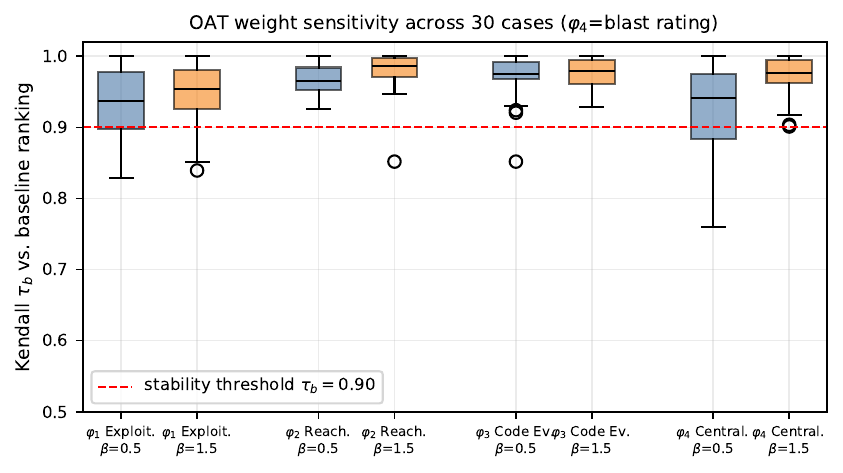}
\caption{OAT weight sensitivity across 30 cases. The median ranking is stable
($\tau_b\!\geq\!0.93$ for every dimension), but \emph{halving} the
exploitability ($\varphi_1$) or centrality ($\varphi_4$) weight destabilises a
sizeable minority of cases (tails reaching $\tau_b\!=\!0.83$ and $0.76$).
Reachability ($\varphi_2$) and code-evidence ($\varphi_3$) are robust to
perturbation.}
\label{fig:oat}
\end{figure}

For each evidence dimension $\varphi_i$ we scale its prior weight
$\bar{w}_i \to \beta\bar{w}_i$ with $\beta\in\{0.5,1.5\}$, renormalise the
remaining weights, re-score, and measure Kendall $\tau_b$ against the
unperturbed ranking. Figure~\ref{fig:oat} reports the distribution across all
30 cases; Table~\ref{tab:oat} summarises it.

\begin{table}[h]
\centering\small
\caption{Per-dimension OAT summary across 30 cases.}
\label{tab:oat}
\begin{tabular}{@{}llcccc@{}}
\toprule
Dim. & $\beta$ & median $\tau_b$ & min $\tau_b$ & mean $\tau_b$ & cases $<\!0.90$ \\
\midrule
$\varphi_1$ Exploit.   & 0.5 & 0.938 & 0.829 & 0.934 & 8 \\
$\varphi_1$ Exploit.   & 1.5 & 0.954 & 0.839 & 0.943 & 6 \\
$\varphi_2$ Reach.     & 0.5 & 0.965 & 0.926 & 0.966 & 0 \\
$\varphi_2$ Reach.     & 1.5 & 0.986 & 0.852 & 0.979 & 1 \\
$\varphi_3$ Code Ev.   & 0.5 & 0.975 & 0.852 & 0.969 & 1 \\
$\varphi_3$ Code Ev.   & 1.5 & 0.980 & 0.928 & 0.977 & 0 \\
$\varphi_4$ Central.   & 0.5 & 0.942 & 0.760 & 0.923 & 10 \\
$\varphi_4$ Central.   & 1.5 & 0.976 & 0.901 & 0.972 & 0 \\
\bottomrule
\end{tabular}
\end{table}

\noindent\textbf{Takeaway.} Sensitivity is asymmetric and concentrated:
\cape's ranking is stable to most weight perturbations, but $\varphi_1$ and
$\varphi_4$ carry the decisive intra-regime ordering signal --- halving either
moves more than a third of cases below the $\tau_b\!=\!0.90$ stability
threshold. This corpus-wide view confirms (and generalises) the single-case
analysis referenced in the paper.

\smallskip\noindent\emph{Where this fits.} This section elaborates
\textbf{RQ4} (weight sensitivity). The paper reports a single representative
sensitivity result; here the same one-at-a-time perturbation is run over all
30 cases so the reader can see the full distribution rather than a point
estimate. It is computed only from the frozen per-case scores and the deployed
AHP prior $\bar{w}$ --- no new tuning. It lets us verify two things the
paper asserts but does not fully visualise: that the ranking is globally stable
to weight noise, and that the stability is \emph{not} uniform --- $\varphi_1$
and $\varphi_4$ are the load-bearing dimensions for intra-regime order. That
asymmetry is the empirical justification for treating those two weights as the
ones worth setting carefully.

\section{Drop-One-Dimension Ablation (30 cases)}
\label{sec:ablation}

\begin{figure}[h]
\centering
\includegraphics[width=0.85\linewidth]{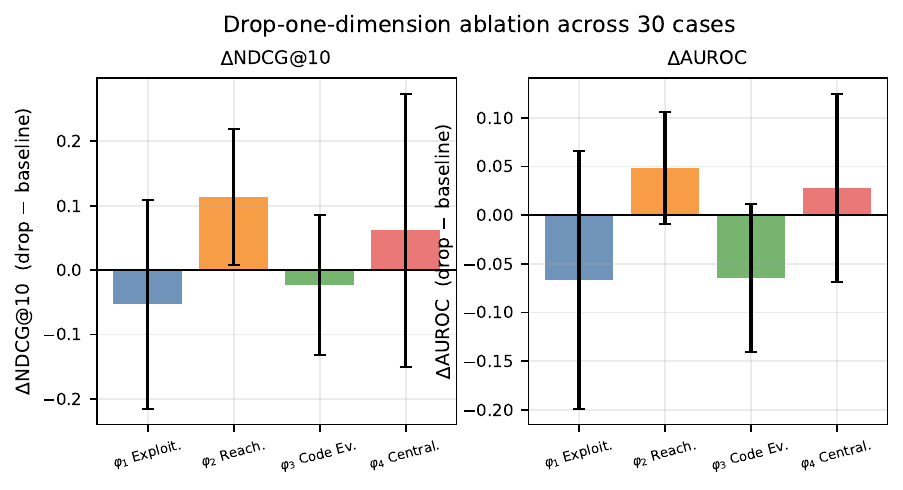}
\caption{Mean $\Delta$NDCG@10 and $\Delta$AUROC when each dimension is dropped
(error bars: population s.d. over 30 cases). Removing exploitability
($\varphi_1$) or code-evidence ($\varphi_3$) degrades both metrics
($\Delta$AUROC $-0.066$ and $-0.065$), confirming they carry non-redundant
ranking signal. Reachability ($\varphi_2$) is largely subsumed by the G2
reachability \emph{gate}, so dropping it as a scoring term is metric-neutral;
the blast-rating $\varphi_4$ is likewise near-neutral on average while still
driving the OAT tail behaviour of \S\ref{sec:oat}.}
\label{fig:ablation}
\end{figure}

We remove each dimension from the active set, renormalise the surviving
weights, and measure the change in ranking quality relative to the full model
(Figure~\ref{fig:ablation}).

\noindent\textbf{Takeaway.} The two evidence dimensions that most distinguish
\cape from severity-only ranking --- proven exploitability and concrete code
evidence --- are also the ones whose removal hurts most. Reachability acts
primarily through the gate rather than the weighted score, which is why it is
robust under OAT yet metric-neutral under ablation.

\smallskip\noindent\emph{Where this fits.} This section also elaborates
\textbf{RQ4}, but from the complementary angle of \emph{necessity} rather than
sensitivity: OAT (\S\ref{sec:oat}) perturbs a weight, whereas this drops a
dimension entirely and re-measures ranking quality ($\Delta$NDCG@10,
$\Delta$AUROC) against the full 4-dimensional model. It is computed by
re-scoring all 30 cases with each $\varphi_i$ removed and the survivors
renormalised. It lets us confirm that every retained dimension earns
its place --- the model is not carrying redundant terms --- and, read together
with Appendix \ref{sec:oat}, it explains the apparent paradox of $\varphi_2$: robust
under weight perturbation yet harmless to drop, because its real work is done by
the G2 gate. This pairing is the evidence behind the paper's claim that the four
dimensions are non-redundant.

\section{Reachability-Penalty ($\alpha$) Sensitivity}
\label{sec:alpha}

\begin{figure}[h]
\centering
\includegraphics[width=0.55\linewidth]{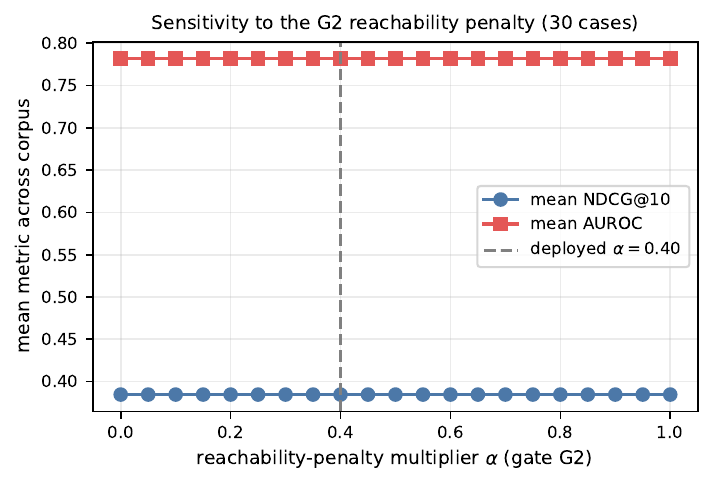}
\caption{Corpus-mean NDCG@10 and AUROC as a function of the G2
reachability-penalty multiplier $\alpha$. Quality rises as unreachable
findings are demoted, plateaus through the deployed value $\alpha\!=\!0.40$,
showing the choice sits on a stable shoulder rather than a sharp optimum.}
\label{fig:alpha}
\end{figure}

The G2 gate multiplies the score of a not-reachable finding in
$\mathcal{R}_3$--$\mathcal{R}_5$ by $\alpha$ (deployed $\alpha\!=\!0.40$).
Figure~\ref{fig:alpha} sweeps $\alpha\in[0,1]$.

\noindent\textbf{What this shows and why it is here.} The deployed score uses a
single hand-set constant, $\alpha\!=\!0.40$, for the G2 reachability penalty:
findings that the call-graph analysis cannot reach in regimes
$\mathcal{R}_3$--$\mathcal{R}_5$ have their score multiplied by $\alpha$ rather
than zeroed. A reviewer could reasonably ask whether that number was reverse-
engineered to flatter the results. This figure answers that: it sweeps the full
$\alpha\in[0,1]$ range and plots corpus-mean ranking quality. The curve rises
monotonically as unreachable findings are demoted (confirming the gate helps at
all), then flattens into a broad plateau that contains $0.40$ --- so quality is
insensitive to the exact value across a wide band, and the deployed choice is
not a fragile optimum. This is a gate-constant robustness check that has no
counterpart in the main paper; it is supplementary-only evidence that the G2
design is sound, not a tuned artefact. (It is one of the two analyses we keep
here rather than fold into \textbf{RQ4}, the other being the grid degeneracy of
\S\ref{sec:grid}.)

\section{Weight Grid Search}
\label{sec:grid}

\begin{figure}[h]
\centering
\includegraphics[width=0.8\linewidth]{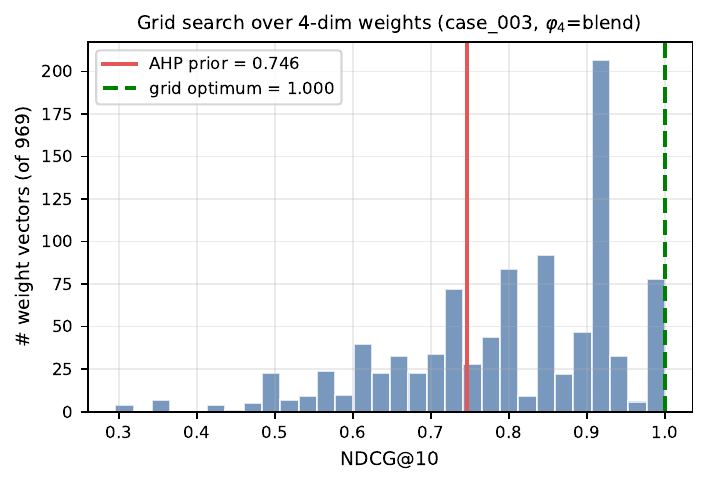}
\caption{NDCG@10 over the full grid of valid 4-dimensional weight vectors
(step 0.05) on the running example (case\_003, $\varphi_4$=blend). The AHP
prior is already near the top of the distribution; the grid optimum is
degenerate (many vectors reach the maximum), so the prior is retained as the
deployment-appropriate, non-overfit choice.}
\label{fig:grid}
\end{figure}

\noindent\textbf{What this shows and why it is here.} This is the exhaustive
counterpart to the local OAT perturbation of \S\ref{sec:oat}: instead of nudging
one weight, it evaluates \emph{every} valid 4-dimensional weight vector on a
0.05 grid and records ranking quality. The point is twofold. First, it locates
the AHP prior $\bar{w}$ within the global landscape --- the prior already sits in
the top band, so \cape is not leaving large amounts of quality on the table by
declining to fit weights per corpus. Second, and more important, the maximum is
\emph{degenerate}: many distinct weight vectors tie for the best score, so there
is no unique data-fitted optimum to chase. Picking one would be overfitting to
this corpus; retaining the theory-derived AHP prior is the honest,
generalisation-safe choice. This elaborates \textbf{RQ4}'s weight-selection
decision, but we deliberately keep the degeneracy argument here rather than in
the main paper --- it justifies \emph{why we did not tune}, which is a
methodological aside, not a headline result.

\section{Ranking-Quality Curves}
\label{sec:curves}

\begin{figure}[h]
\centering
\includegraphics[width=0.92\linewidth]{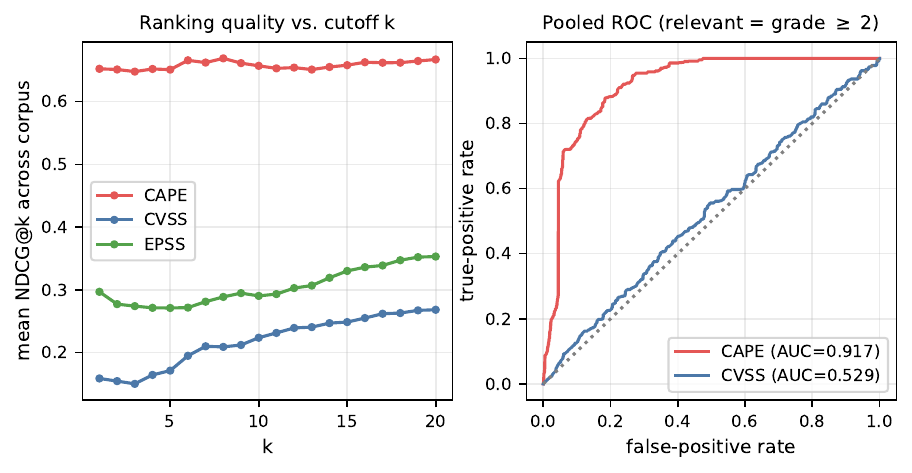}
\caption{Left: mean NDCG@$k$ across the corpus as the cutoff $k$ varies; \cape
dominates CVSS and EPSS at every depth. Right: pooled ROC over all findings
(relevant $=$ ground-truth grade $\geq 2$); \cape separates relevant from
irrelevant findings far better than CVSS.}
\label{fig:curves}
\end{figure}

\noindent\textbf{What this shows and why it is here.} This section elaborates
\textbf{RQ3} (ranking quality). The paper reports NDCG@10 and AUROC as single
numbers at a fixed cutoff; those numbers are vulnerable to the objection that
\cape was simply tuned to look good at $k\!=\!10$. The two curves remove that
doubt. The left panel sweeps the cutoff $k$ and shows \cape dominating both CVSS
and EPSS at \emph{every} depth, not just the reported one --- the advantage is a
property of the ranking, not of the chosen $k$. The right panel pools all
findings across all cases into a single ROC (relevant $=$ ground-truth grade
$\geq 2$), giving a threshold-independent view of how well the score separates
exploitable-and-relevant findings from the rest. Both are computed from the same
frozen ranked outputs that produce the paper's RQ3 table; they add the shape of
the curve to the point estimate, letting a reader confirm the gain is broad and
monotone rather than a single fortunate operating point.

\section{Centrality-Variant Study}
\label{sec:centrality}

\begin{figure}[h]
\centering
\begin{minipage}{0.46\linewidth}
\centering
\includegraphics[width=\linewidth]{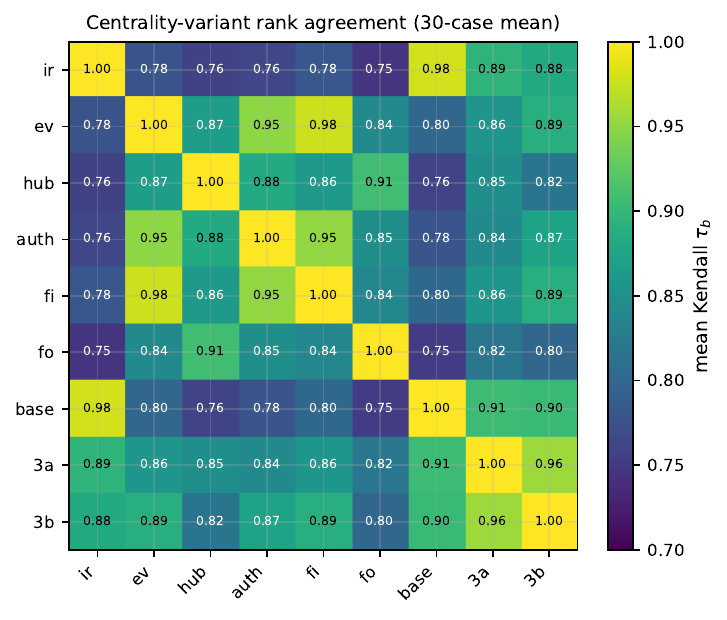}
\caption{Mean pairwise Kendall $\tau_b$ between centrality variants (30-case
mean). \texttt{ev}/\texttt{auth}/\texttt{fi} form a tight cluster; \texttt{hub}
is the most distinct, which is why it supplies the complementary signal in
Selection\_3a.}
\label{fig:tau}
\end{minipage}\hfill
\begin{minipage}{0.50\linewidth}
\centering
\includegraphics[width=\linewidth]{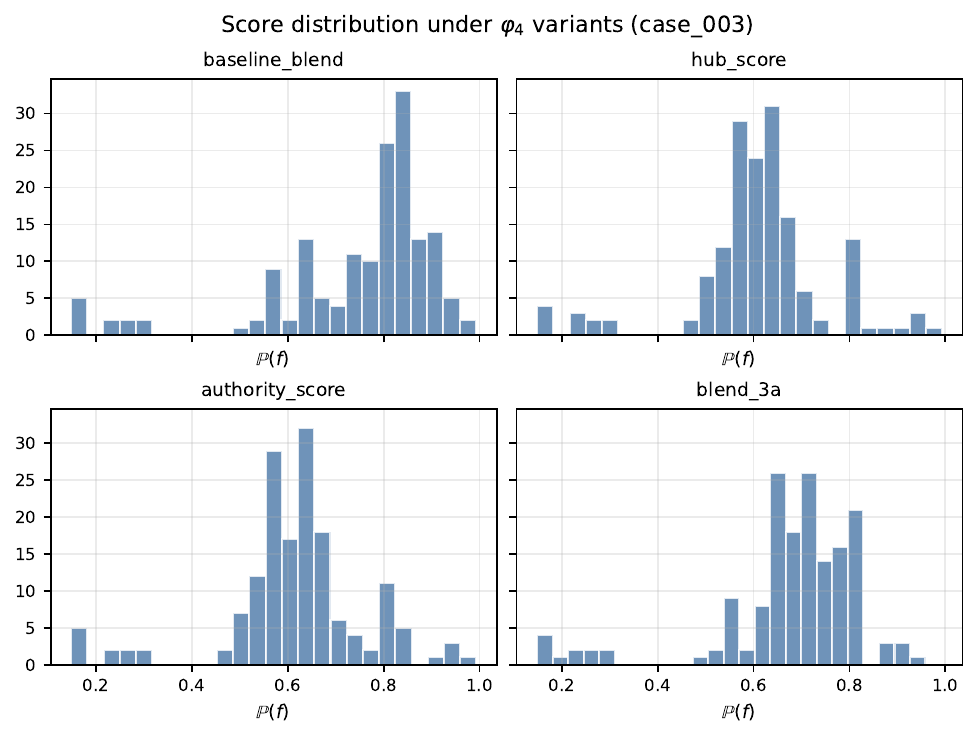}
\caption{Distribution of final priority $\mathbb{P}(f)$ on case\_003 under four
$\varphi_4$ choices. The blend (\texttt{blend\_3a}) spreads scores more
smoothly than single-metric variants, aiding fine-grained ranking.}
\label{fig:pfinal}
\end{minipage}
\end{figure}

Figure~\ref{fig:tau} shows the rank agreement between centrality realisations
of $\varphi_4$; Figure~\ref{fig:pfinal} shows how the score distribution
changes with the choice.

\noindent\textbf{What this shows and why it is here.} This section elaborates
the \emph{centrality-metric selection} half of \textbf{RQ4}. The centrality
dimension $\varphi_4$ can be realised by several graph metrics (eigenvector
\texttt{ev}, authority \texttt{auth}, fan-in \texttt{fi}, hub \texttt{hub},
$\dots$), and the paper selects a blend (Selection\_3a). These two figures
justify that choice. The left heatmap (Figure~\ref{fig:tau}) reports the mean
pairwise Kendall $\tau_b$ between the rankings each metric induces, averaged over
30 cases: \texttt{ev}/\texttt{auth}/\texttt{fi} agree closely (they are
near-redundant), while \texttt{hub} is the outlier --- it carries
\emph{different} structural information, which is exactly why blending it in adds
signal a single dominant metric would miss. The right histogram
(Figure~\ref{fig:pfinal}) shows the downstream effect on case\_003: the blend
spreads final priority $\mathbb{P}(f)$ more smoothly than any single-metric
variant, which reduces ties and sharpens fine-grained ordering. Together they
let a reader verify that the blend is a principled choice (distinct-then-combine)
rather than an arbitrary one, supplying the agreement evidence behind the RQ4
selection narrative.

\section{Evidence Dimensions on the Running Example}
\label{sec:anatomy}

\begin{figure}[h]
\centering
\begin{minipage}{0.36\linewidth}
\centering
\includegraphics[width=\linewidth]{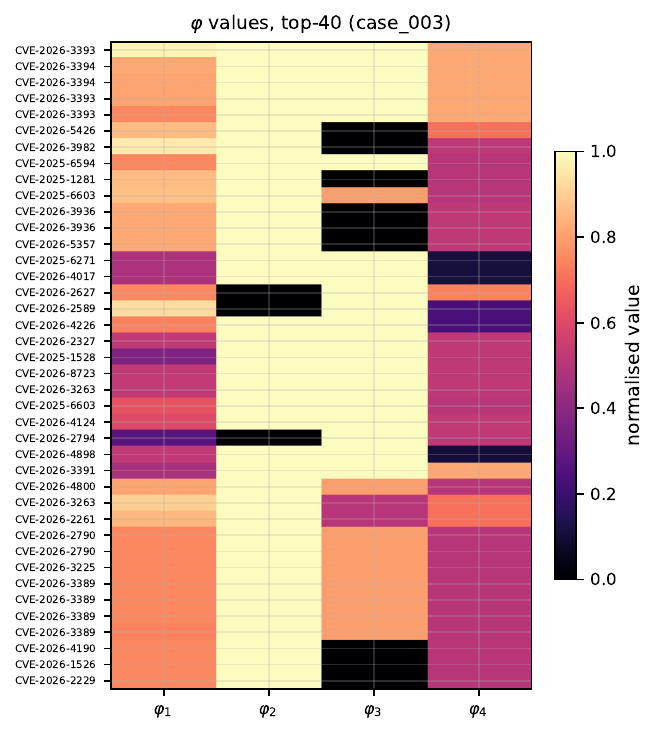}
\caption{Normalised $\varphi_1$--$\varphi_4$ values for the top-40 findings of
case\_003. The dependency-type dimension is excluded from the deployed
4-dimensional model.}
\label{fig:phi}
\end{minipage}\hfill
\begin{minipage}{0.60\linewidth}
\centering
\includegraphics[width=\linewidth]{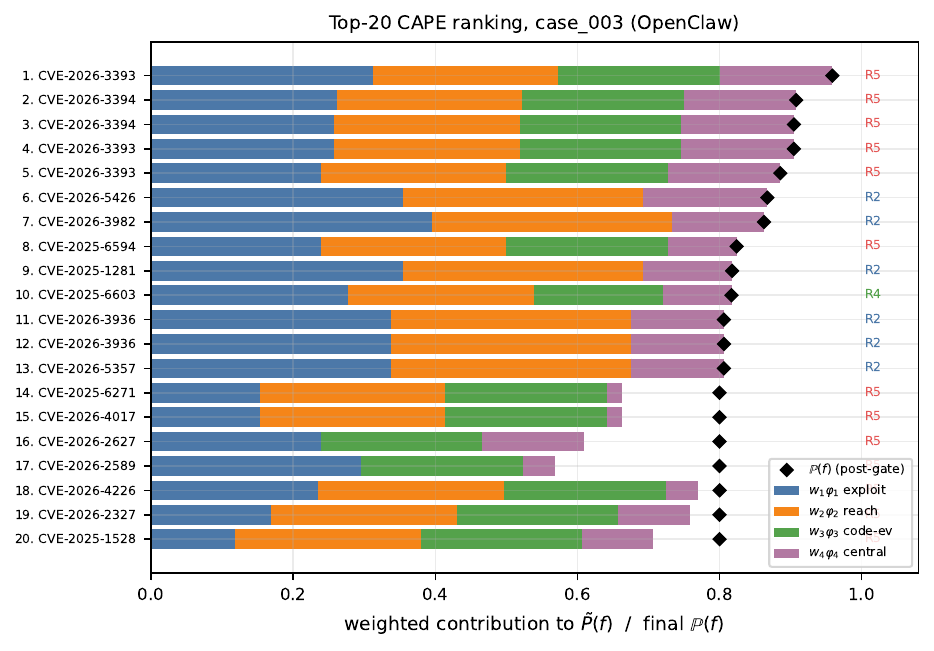}
\caption{Anatomy of the top-20 \cape ranking on case\_003. Bars decompose
$\tilde{P}(f)$ into the weighted contribution of each dimension; black
diamonds mark the gate-adjusted final score $\mathbb{P}(f)$ (e.g.\ the
$\mathcal{R}_5$ floor lifting proven-exploitable findings). Regime labels at
right.}
\label{fig:anatomy}
\end{minipage}
\end{figure}

\noindent\textbf{What this shows and why it is here.} This pair opens up the
scoring model $\mathbb{P}(f)$ itself on case\_003, the running example used
throughout the paper --- it elaborates the priority-score mechanism rather than
any single RQ. The left heatmap (Figure~\ref{fig:phi}) shows the normalised
$\varphi_1$--$\varphi_4$ inputs for the top-40 findings, so a reader can see
\emph{which} evidence each finding actually carries before any weighting. The
right bar chart (Figure~\ref{fig:anatomy}) then decomposes the pre-gate weighted
score $\tilde{P}(f)$ into per-dimension contributions and overlays the
gate-adjusted final $\mathbb{P}(f)$ as black diamonds, with regime labels. This
makes the model auditable end-to-end: one can trace why a given finding ranks
where it does --- which dimension dominated, and how a gate (e.g.\ the
$\mathcal{R}_5$ floor lifting a proven-exploitable finding, or the G2 penalty
demoting an unreachable one) moved it. It adds a worked, per-finding explanation
that the main paper has no room for, turning the score from a black box into an
inspectable computation on a concrete case.

\begin{figure}[h]
\centering
\includegraphics[width=0.95\linewidth]{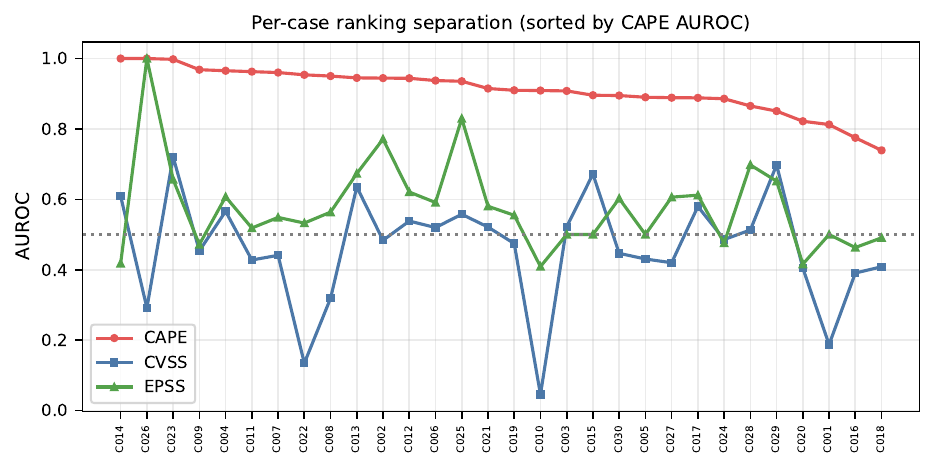}
\caption{Per-case AUROC for \cape, CVSS, and EPSS across all 30 cases (sorted
by \cape AUROC). \cape is above the random-baseline line (0.5) in every case
and dominates both baselines throughout.}
\label{fig:auroc}
\end{figure}

\noindent\textbf{What this shows and why it is here.} Figure~\ref{fig:auroc}
returns to \textbf{RQ3} but reports it per case rather than pooled: one AUROC bar
per case for \cape, CVSS, and EPSS, sorted by \cape's score. Where the curves of
\S\ref{sec:curves} show the corpus-average advantage, this shows its
\emph{consistency} --- \cape stays above the 0.5 random baseline in all 30 cases
and above both baselines throughout, so the headline RQ3 gain is not an average
hiding a few catastrophic cases. It lets a sceptical reader check the worst case,
not just the mean, and confirms the ranking advantage is corpus-wide rather than
driven by a handful of favourable repositories.

\end{document}